\documentclass[a4paper,11pt]{article}
\usepackage[left=1in,right=1in,top=1in,bottom=1.5in]{geometry}
\usepackage{amsmath}
\usepackage{amsthm}
\usepackage{array}
\usepackage{amsfonts}
\usepackage{amssymb}
\usepackage{amsopn}
\usepackage{graphicx}
\usepackage{authblk}
\usepackage[usenames,dvipsnames]{xcolor}
\usepackage[numbers]{natbib} 

\usepackage{hyperref}
\usepackage{ifpdf}

\ifpdf

\else

\def\longrightharpoonup{\relbar\joinrel\rightharpoonup}
\def\longleftharpoondown{\leftharpoondown\joinrel\relbar}

\def\longrightleftharpoons{
  \mathop{
    \vcenter{
      \hbox{
      \ooalign{
        \raise1pt\hbox{$\longrightharpoonup\joinrel$}\crcr
	  \lower1pt\hbox{$\longleftharpoondown\joinrel$}
	  }
      }
    }
  }
}

\newcommand{\rates}[2]{\displaystyle
\mathrel{\longrightleftharpoons^{#1\mathstrut}_{#2}}}
\newcommand{\br}[1]{\langle #1\rangle}

\title{\bf Stochastic focusing coupled with negative feedback enables robust regulation in biochemical reaction networks}
\author[1]{Andreas Milias-Argeitis}
\author[2]{Stefan Engblom}
\author[2]{Pavol Bauer}
\author[1]{Mustafa Khammash\thanks{mustafa.khammash@bsse.ethz.ch}}
\affil[1]{Department of Biosystems Science and Engineering, ETH Zurich, Mattenstrasse 26, 4058 Basel, Switzerland}
\affil[2]{Division of Scientific Computing, Department of Information Technology, Uppsala University, P.O. Box 337, SE-75105 Uppsala, Sweden}

\date{}

\begin{document}

\maketitle

\begin{abstract}

  Nature presents multiple intriguing examples of processes which
  proceed at high precision and regularity.  This remarkable stability
  is frequently counter to modelers' experience with the inherent
  stochasticity of chemical reactions in the regime of low copy
  numbers.  Moreover, the effects of noise and nonlinearities can lead
  to ``counter-intuitive'' behavior, as demonstrated for a basic
  enzymatic reaction scheme that can display \emph{stochastic
    focusing} (SF). Under the assumption of rapid signal fluctuations,
  SF has been shown to convert a graded response into a threshold
  mechanism, thus attenuating the detrimental effects of signal
  noise. However, when the rapid fluctuation assumption is violated,
  this gain in sensitivity is generally obtained at the cost of very
  large product variance, and this unpredictable behavior may be one
  possible explanation of why, more than a decade after its
  introduction, SF has still not been observed in real biochemical
  systems.

  In this work we explore the noise properties of a simple enzymatic
  reaction mechanism with a small and fluctuating number of active
  enzymes that behaves as a high-gain, noisy amplifier due to SF
  caused by slow enzyme fluctuations.  We then show that the inclusion
  of a plausible negative feedback mechanism turns the system from a
  noisy signal detector to a strong homeostatic mechanism by
  exchanging high gain with strong attenuation in output noise and
  robustness to parameter variations. Moreover, we observe that the
  discrepancy between deterministic and stochastic descriptions of
  stochastically focused systems in the evolution of the means almost
  completely disappears, despite very low molecule counts and the
  additional nonlinearity due to feedback.

  The reaction mechanism considered here can provide a possible
  resolution to the apparent conflict between intrinsic noise and high
  precision in critical intracellular processes.



\end{abstract}


\section{Introduction}

Random fluctuations due to low-copy number phenomena inside the
microscopic cellular volumes have been an object of intense study in
recent years. It is now widely recognized that deterministic modeling
of chemical kinetics is in many cases inadequate for capturing even
the mean behavior of stochastic chemical reaction networks, and
several studies have explored the discrepancy between deterministic
and stochastic system descriptions \cite{circadian,Paulsson00,
  newsteadystates_RDME,Samoilov05}.

Despite the all-pervasive stochasticity, cellular processes and
responses proceed with surprising precision and regularity, thanks to
efficient noise suppression mechanisms also present within cells. The
structure and function of these mechanisms has been a topic of great
interest \cite{ElSamad06,Becskei00,Thattai02,Swain04,Hornung08,
  Osella11,Burger12,Samoilov05}, and in many cases still remains
unknown.

Moreover, recent theoretical works on enzymatic reaction schemes with
a single or a few enzyme molecules \cite{Kou05,English05,Grima14,
  Schwabe13} have repeatedly shown that low-copy enzymatic reactions
demonstrate a stochastic behavior that can lead to markedly different
responses in comparison to the predictions of deterministic enzyme
kinetic models.

In this work we investigate the properties of a possible noise
suppression mechanism for an enzymatic reaction with a small and
fluctuating number of active enzymes. Under certain conditions,
presented in \cite{Paulsson00}, this system displays an increased
sensitivity to enzyme fluctuations, a phenomenon that has been termed
\emph{stochastic focusing}.

Stochastic focusing has been presented as a possible mechanism for
\emph{sensitivity amplification}: compared to a deterministic model of
a biochemical network, the mean output of the stochastic version of
the system can display increased sensitivity to changes in the input,
when the input species has sufficiently low abundance. Consequently,
it has been postulated that stochastic focusing can act as a signal
detection mechanism, that converts a graded input into a ``digital''
output.

The basic premise of \cite{Paulsson00} has been that fluctuations in
the ``input'' species are sufficiently rapid, so that any rates that
depend on the signaling species show minimal time-correlations. We
show that if this condition fails, i.e.~when the fluctuations in the
input signal are slow compared to the average lifetime of a substrate
molecule, stochastic focusing can result in a dramatic increase in
substrate fluctuations, a fact also acknowledged in the original
publication. Increased sensitivity to input changes does not only come
at the cost of extremely high output noise levels; as we will
demonstrate here, systems operating in this regime are also extremely
sensitive to variations in reaction rates, which in fact precludes
robust signal detection by stochastic focusing.

For the first time since its introduction we could study the
steady-state behavior of this system analytically, by formulating and
solving the equations for the conditional means and (co)variances
\cite{Hasenauer13}. Motivated by our observations on the open-loop,
stochastically focused system, we investigated the system behavior in
the presence of a plausible feedback mechanism. We treated the enzyme
as a noisy ``controller'' molecule whose purpose is to regulate the
outflux of a reaction product by -- directly or indirectly --
``sensing'' the fluctuations in its substrate. For the sake of
simplicity and clarity, we focused on very simple and highly
abstracted mechanisms, but we should remark that several possible
biochemical implementations of our feedback mechanism can be
considered.

Our premise was that the great open-loop sensitivity of a
stochastically focused system with relatively slow input fluctuations
creates a system with very high open-loop gain, which in turn can be
exploited to generate a very robust closed-loop system once the output
is connected to the input. Our simulation results confirmed this
intuition, revealing a dramatic decrease in noise levels and a
significant increase in robustness in the steady-state mean behavior
of the closed-loop system. Such a system no longer functions as a
signal detector, but rather behaves as a strong homeostatic mechanism.
Moreover, we observed that the steady-state behavior of the means in a
stochastically focused system with feedback can be captured quite
accurately by the corresponding deterministic system of
  reaction rate equations, despite the fact that the stochastic system
  still operates at very low copy numbers.

Noise attenuation through feedback and the fundamental limits of any
feedback system implemented with noisy ``sensors'' and ``controllers''
have been studied theoretically in the recent years, and some fairly
general performance bounds have been derived in \cite{Paulsson10}.  We
should note that, despite its generality, the modeling framework
assumed in \cite{Paulsson10} does not apply in our case, since our
system contains a controlled degradation reaction, whereas
\cite{Paulsson10} considers only control of production.
  More specifically, \cite{Paulsson10} examines the case where a given
  species regulates its own production through an arbitrary stochastic
  signaling network. In this setting, it is shown that, no matter the
  form or complexity of the intermediate signaling, the loss of
  information induced by stochasticity places severe fundamental
  limits on the levels of noise suppression that such feedback loops
  can achieve. On the other hand, it is still unclear what type of
noise suppression limitations are present for systems such as the one
studied here, and a complete analytical treatment of the problem of
regulated substrate degradation seems very difficult at the moment.


A first attempt to analyze the noise properties of regulated
degradation was presented in \cite{ElSamad06}, which examined such a
scheme using the Linear Noise Approximation (LNA) \cite{Elf03}. As the
authors of that work pointed out, however, the LNA is incapable of
correctly capturing the system behavior (i.e. means and variances)
beyond the small-noise regime, due to the nonlinear system
behavior. We verified this inadequacy, not only for LNA, but for other
approximation schemes as well, such as the Langevin equations
\cite{Mazza14} and various moment closure approaches
\cite{Singh11}. Perhaps this is the reason why, contrary to regulated
production, the theoretical noise properties of regulated substrate
degradation have received relatively little attention.

With the rapid advancement of single-molecule enzymatic assays
\cite{Sauer10,Mashanov11}, we expect that the study of noise
properties of various low-copy enzymatic reactions, including the
proposed feedback mechanism described here, will soon be amenable to
experimental verification. It also remains to be seen whether the
proposed feedback mechanism is actually employed by cells to achieve
noise attenuation in enzymatic reactions. At any rate, the noise
attenuation scheme presented here could be tried and tested in
synthetic circuits through enzyme engineering \cite{Raman14,Cross13}.



\section{Results}
\subsection{A slowly fluctuating enzymatic reaction scheme that exhibits stochastic focusing}

In this section we formulate and analyze a simple biochemical reaction
network capable of exhibiting the dynamic phenomenon of stochastic
focusing. It is shown that in the stochastic focusing regime, the
system acts as a noisy amplifier with an inherent strong sensitivity
to perturbations. It follows that without modifications, the network
cannot be used under conditions requiring precision and regularity.

\subsubsection{Modeling}

\begin{figure}[h!tb]
\centering
  \includegraphics[width=\textwidth]{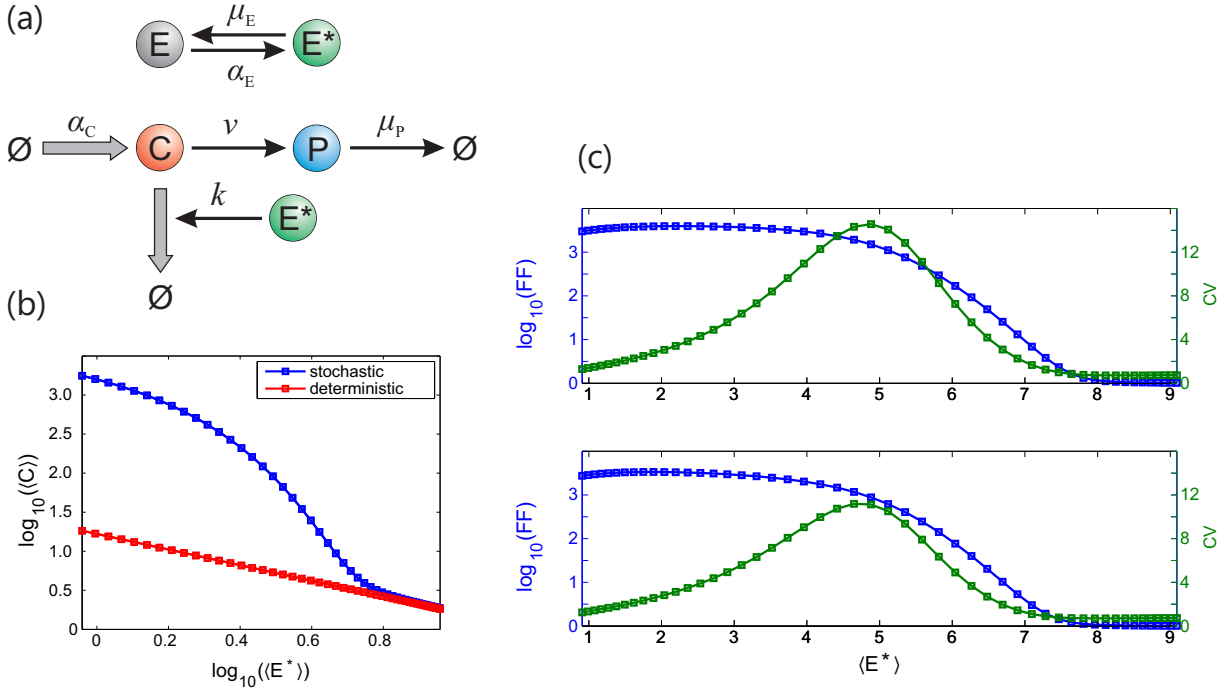}
  \caption{{\bf (a)} The reaction scheme studied in this work. Thick
    arrows represent high-flux reactions. {\bf (b)} Blue line:
    Steady-state mean of $C$ (denoted by $\br{C}$ throughout) as a
    function of the average number of active enzyme molecules
    ($N\alpha_E/(\alpha_E+\mu_E))$ for
    $N=10,~\mu_E=0.1,~\alpha_C=5010,~k=300,~\nu=1,~\mu_P=1$. Since
    $\nu=\mu_P$, the average of $P$ displays the same behavior as the
    substrate. Red line: Steady-state of the ODE model for the same
    parameter values. The large difference (notice the logarithmic
    scale) between the blue and red lines is a consequence of
    stochastic focusing \cite{Paulsson00}. {\bf (c)} Upper: Substrate
    ($C$) noise statistics as a function of the average number of
    active enzymes. Blue: Steady-state Fano factor (variance/mean;
    notice the logarithmic scale). Green: Steady-state coefficient of
    variation (standard deviation/mean). Lower: Product ($P$) noise
    statistics as a function of the average number of active
    enzymes. Color coding same as in upper panel. Both plots were
    obtained for
    $N=10,~\mu_E=0.1,~\alpha_C=5010,~k=300,~\nu=1,~\mu_P=1$. All
    calculations were performed analytically, using the conditional
    moment equations and the known stationary distribution of $E^*$.}
  \label{Fig1_OL}
\end{figure}

We consider the simple branched reaction scheme studied in
\cite{Paulsson00} and shown schematically in Fig.~\ref{Fig1_OL}(a). In
this scheme, substrate molecules $C$ enter the system at a constant
influx, and can either be converted into a product $P$ or degraded
under the action of a low-copy enzyme (or, equivalently, converted
into a product that leaves the system). While the number of enzymes in
the system is assumed constant, enzyme molecules can spontaneously
fluctuate between an active ($E^*$) and an inactive ($E$) form.  The
generality of this model and its sensitivity to variations in the
active enzyme levels is further discussed in the Supplement (sections 1 and 2).

Recent single-enzyme turnover experiments have shown that single
enzyme molecules typically fluctuate between conformations with
different catalytic activities, a phenomenon called \emph{dynamic
  disorder} \cite{Kou05,English05,Grima14,Schwabe13}.  In the simple
model considered here, the enzyme randomly switches between two
activity states.  The stationary distribution of $E^*$ in this case is
known to be binomial \cite{Ullah11}; that is, $E^*\sim\mbox{B}(N,p)$,
where $N = E+E^*$ is the total number of enzymes in the system and $p
= \alpha_E/(\alpha_E+\mu_E)$, such that the mean $\langle E^* \rangle
= Np$.

The basic (empirically derived) conditions for stochastic focusing
\cite{Paulsson00} are that the magnitude of active enzyme fluctuations
is significant compared to the mean number of active enzymes, while
the total number of enzymes is low. Moreover, it is assumed that the
level of $E^*$ fluctuates rapidly compared to the average lifetime of
$C$ and $P$ molecules. Without this assumption, the noise in $E^*$ can
be greatly amplified by $C$ and transmitted to $P$.

The first assumption (large enzyme fluctuations and low abundance) is
maintained in our setup. However, we shall dispense with the second
assumption. We further postulate that $C$ (possibly a product of
upstream enzymatic reactions) enters the system at a high input flux
(large $\alpha_C$) and that there exists a strong coupling between $C$
and $E^*$, in the sense that a few active enzymes can strongly affect
the degradation of $C$.

More concretely, the previously stated assumptions imply that the
reaction rates must satisfy the following conditions:
\begin{enumerate}
\item $\alpha_C\gg 1$ (high influx of $C$)
\item $k\gg \alpha_E+\mu_E$ (enzyme fluctuations are slow compared to
  the average lifetime of a substrate molecule)
\item $k\gg 1$ (strong coupling between enzyme and substrate)
\item $Np = N \displaystyle\frac{\alpha_E}{\alpha_E+\mu_E}$ is small
  (e.g.~below 10).
\end{enumerate}
These conditions are motivated via a short theoretical and numerical
analysis in the Supplement (section 1). When they hold, we expect the amount of
$C$ to fluctuate wildly as $E^*$ varies over time and these
fluctuations to propagate to $P$. In the rest, we will refer to this
motif as the (open-loop) \emph{slowly fluctuating enzyme} (SFE)
system.

The computational analysis of this and similar systems has thus far
been hindered by the presence of the bimolecular reaction, which leads
to statistical moment equations that are not closed \cite{Singh11},
while the presence of stochastic focusing presents further
difficulties for any moment closure method. In this work, we
circumvent these difficulties by formulating and solving the
conditional moment equations \cite{Hasenauer13} for the means and
(co)variances of $C$ and $P$ conditioned on the enzyme state (whose
steady-state distribution is known). This enables for the first time
the analytical study of the steady-state behavior of this system (more
details can be found in the Supplement, section 3). Chiefly, the equations for
the first two conditional moments of the SFE system are in fact
closed, i.e.~they do not depend on moments of order higher than 2, and
thus do not require a moment closure approximation despite the fact
that the unconditional moment equations themselves are open. We next
use these analytic equations to shed new light on the properties of
the network under consideration.

\subsubsection{The SFE system functions as a noisy amplifier}
\label{subsubsec:pert_open}

According to the method of conditional moments (MCM)
  \cite{Hasenauer13}, the chemical species of a given system are
  divided into two classes. Species of the first class, collectively
  denoted by $Y$, are treated fully stochastically, while species of
  the second class, denoted by $Z$, are described through their
  conditional moments given $Y$.  More analytically, the MCM considers
  a chemical master equation (CME) \cite{VanKampen92} for the marginal
  distribution of $Y$, $p(Y,t)=\sum_{Z\geq 0}p(Y,Z,t)$, and a system
  of conditional means ($\mu_Z(Y,t):=\mathbb{E}_{p(Z|Y,t)}[Z|Y,t]$)
  and higher-ordered centered moments (e.g. conditional variances
  $C_Z(Y,t):=\mathbb{E}_{p(Z|Y,t)}[(Z-\mu_Z(Y,t))^2|Y,t]$) for the $Z$
  species. In the case of the SFE network, by taking $Y=E^*$ and
  $Z=(C,P)$, we see that $Y$ is independent of $Z$ and its evolution
  is described by a CME whose stationary solution is known. Moreover,
  the system of conditional means and (co)variances of $Z$ given $Y$
  turns out to be closed.

We thus begin by examining the behavior of the system as the enzyme
activation rate is varied while keeping other parameters
unchanged. Assuming a fixed $\mu_E$, $\alpha_E$ is
  directly related to $\langle E^*\rangle$, the average number of
  active enzymes. Thus, any changes in $\langle E^*\rangle$ are
  assumed to be driven by $\alpha_E$, which means that the two can be
  used interchangeably. We present the performance of the open-loop
  system with respect to $\langle E^*\rangle$ wherever possible, as we
  find this more intuitive.

The results in Fig.~\ref{Fig1_OL}(b) show that the stationary means of
$C$ and $P$ (denoted by angle brackets throughout the paper) depend
very sensitively on $\langle E^*\rangle$, as one would expect from a
stochastically focused system. Moreover, owing to the relatively slow
switching frequency of enzyme states, the stationary distributions of
substrate and product are greatly over-dispersed, as shown in
Fig.~\ref{Fig1_OL}(c).



\begin{figure}[h!tb]
\centering
  \includegraphics[width=\textwidth]{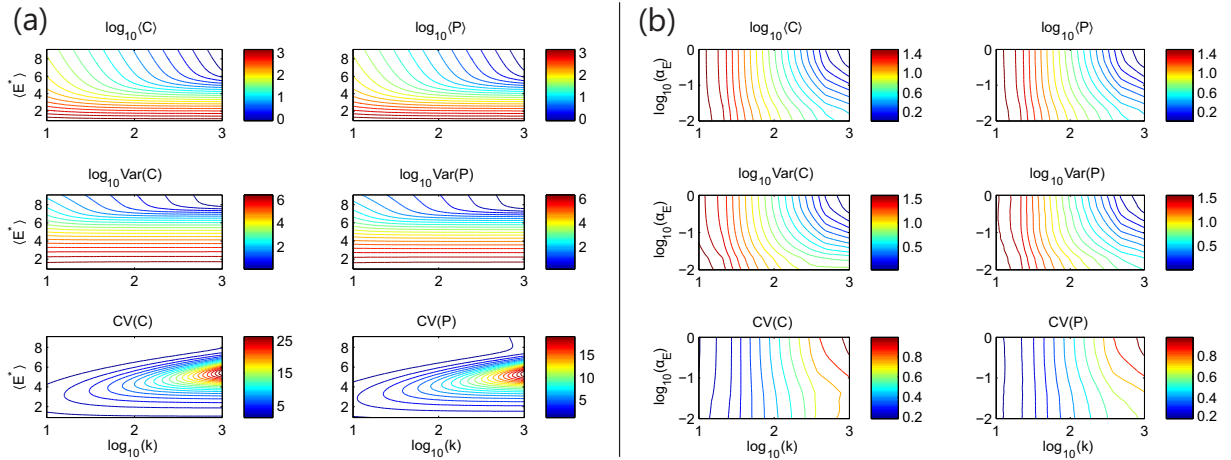}
  \caption{\textbf{(a)} Steady-state means, variances and CVs of
    substrate and product as a function of the average active enzymes
    ($\alpha_EN/(\alpha_E+\mu_E)$) and $k$ for
    $N=10,~\mu_E=0.1,~\alpha_C=5010,~\nu=1,~\mu_P=1$ (notice the
    logarithmic scales). \textbf{(b)} Closed-loop SFE system with
    substrate feedback: steady-state means, variances and CVs of
    substrate and product as a function of $\alpha_E$ (which
    determines the average number of active enzymes) and $k$ for
    $K=3,~C_0=5,~N=10,~\mu_E=0.1,~\alpha_C=5010,~\nu=1,~\mu_P=1$.
    Logarithmic scales are preserved to make comparisons with
    Fig.~\ref{2dscan} easier, although the range of variation is much
    smaller in this case.}
  \label{2dscan}
\end{figure}

Apart from the enzyme activation rate, the catalytic degradation rate
($k$) is also expected to affect noise in the system, as it controls
both the timescale and magnitude of substrate fluctuations: as $k$
increases, the rate of substrate consumption grows as well. On the
other hand, the impact of a change in the number of active enzymes is
also magnified. We can study the interplay of $\alpha_E$ and $k$ by
varying both simultaneously, as shown in Fig.~\ref{2dscan}(a). Although
$\alpha_E$ has a much more pronounced effect on substrate and product
means and variances, the interplay of $k$ and $\alpha_E$ is what
determines the overall noise strength in the system, as the third row
of plots shows.

From the above analysis, we deduce that the open-loop
  motif amplifies both small changes in the average number of active
  enzymes (Fig.~\ref{Fig1_OL}(b)), as well as temporal fluctuations in
  the active enzyme levels (Fig.~\ref{Fig1_OL}(c)): for intermediate
  values of $\alpha_E$, the CV and FF of $C$ and $P$ are much greater
  than zero. This implies that the instantaneous flux of substrate
  through the two alternative pathways experiences very large
  fluctuations, which would propagate to any reactions downstream of
  $C$.

\subsubsection{The SFE system is very sensitive to parameter perturbations}
\label{sub_pert_open}

\begin{figure}[h!tb]
\centering
  \includegraphics[width=0.8\textwidth]{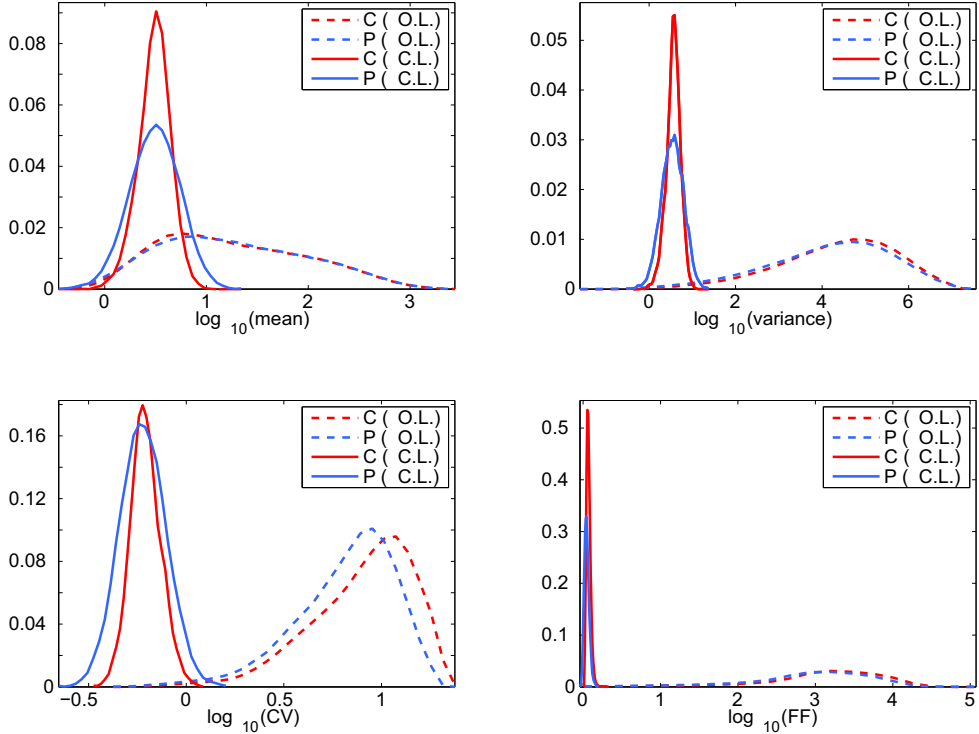}
  \caption{{\bf Dashed lines}: Histograms of steady-state means,
    variances, CVs and Fano Factors of substrate and product, obtained
    from 10000 randomly sampled parameters, following
    \eqref{pert}. Nominal values of perturbed parameters:
    $N=10,~\alpha_E=0.08,~\mu_E=0.1,~\alpha_C=5010,~k=300,~\nu=1,~\mu_P=1$.
    The black line on the top left plot denotes the (common) mean of
    substrate and product for the nominal parameters. On the top right
    plot, black lines mark the nominal variance for substrate (solid)
    and product (dashed). {\bf Continuous lines}: Closed-loop SFE
    system ($K=3,~C_0=5$) with substrate feedback:
      Histograms of steady-state means, variances, CVs and FFs of
      substrate and product}, obtained from the same 10000 randomly
    sampled parameters used for the dashed line histograms. The great
    reduction in sensitivity of the closed-loop SFE system in
    comparison to the open-loop can be easily observed.
  \label{fig:pert_open_closed}
\end{figure}

The increased sensitivity of the SFE network to fluctuations in the
active enzyme would suggest sensitivity with respect to variations in
reaction rates. To verify this, we generated 10000 uniformly
distributed joint random perturbations of all system parameters that
reach up to 50\% of their nominal values. That is, every parameter was
perturbed according to the following scheme:
\begin{equation}
  \label{pert}
  p_{pert} = p_{nom}+(n-0.5)p_{nom},~n\sim\mathcal{U}([0,1]).
\end{equation}
For each perturbed parameter set, the steady-state conditional moment
equations were solved to obtain the means, variances and noise
measures for both the substrate and product. The results are
summarized in Fig.~\ref{fig:pert_open_closed} (dashed lines), where
the large parametric sensitivity of the system can be clearly
seen. Global sensitivity analysis of the mean, variance and CV
histograms \cite{Saltelli08} reveals that the total number of enzymes
($N$) has the largest effect on all these quantities, with the enzyme
activation/deactivation rates ($\alpha_E,~\mu_E$) coming at second and
third place. Although one could argue that $\alpha_E$ and $\mu_E$ are
biochemical rates that are uniquely determined by molecular features
of the enzyme, the total number of enzyme molecules would certainly be
variable across a cellular population.

Taken together, the results of this section and the previous one
suggest that the operation of the SFE reaction scheme in
Fig.~\ref{Fig1_OL}(a) as a signal detection mechanism (the
original point made in \cite{Paulsson00}) is severely compromised when
the system operates in the regime defined by our set of assumptions:
besides amplifying enzyme fluctuations, the system responds very
sensitively to parametric perturbations. These features render the
enzyme a highly non-robust controller of the substrate and product
outfluxes, which can fluctuate dramatically in time. In addition,
reaction rates have to be very finely tuned to achieve a certain
output behavior, for example a given mean and variance, or a given
average substrate outflux.



\subsection{Closing the loop: The SFE network with negative feedback}

It is a well-known fact in control theory that negative feedback
results in a reduction of the closed-loop system gain
\cite{Dorf11,khammashBMC}. However, this reduction is exchanged for
increased stability and robustness to input fluctuations, and a more
predictable system behavior that is less dependent on parameter
variations. Systems with large open-loop gain tend to also display
extreme sensitivity to input and parametric perturbations, and can
thus benefit the most from the application of negative feedback.

We shall examine the operation of the SFE network under feedback by
assuming that $C$ (or $P$) affects the rate of activation of the
enzyme, for example by controlling its activation rate. We will call
this new motif the \emph{closed-loop} SFE system, to differentiate it
from the open-loop system presented above.

\begin{figure}[h!tb]
\centering
  \includegraphics[width=0.9\textwidth]{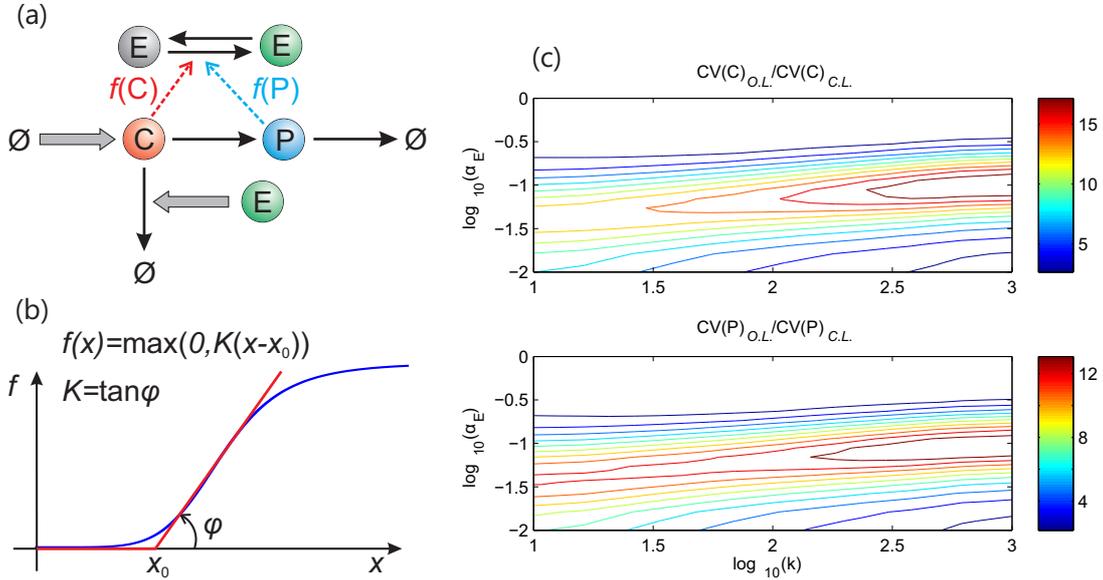}
  \caption{\textbf{(a)} The SFE network with feedback (two possible
    configurations). \textbf{(b)} Red line: the class of feedback
    functions considered in this work, which can approximate a Hill
    function around its lower end. \textbf{(c)} Ratio of open-loop
    vs.~closed-loop CVs as a function of $\alpha_E$ (which determines
    the average number of active enzymes) and $k$ for
    $K=3,~C_0=5,~N=10,~\mu_E=0.1,~\alpha_C=5010,~\nu=1,~\mu_P=1$.}
  \label{closed_scheme}
\end{figure}

According to the closed-loop reaction scheme
(Fig.~\ref{closed_scheme}(a)), the activation rate of $E$ becomes
$\alpha_E(1+f(x))$ ($x$ being $C$ or $P$), where $f$ models activation
by $x$, thus creating a negative feedback loop between the system
input and output. Our only requirement for $f$ is to be nondecreasing
(e.g.~a Hill function). To facilitate our simulation-based analysis,
we will assume that $f$ arises from the local, piecewise linear
approximation of a Hill function, as shown in
Fig.~\ref{closed_scheme}(b). In this case, the form of $f$ is
controlled by two parameters: $K$ (the ``gain'') and $x_0$ (the point
beyond which feedback is activated). Finally, $\alpha_E$ can be
thought of as the ``basal'' activation rate in the absence of the
regulating molecule.

We should note that the proposed form of feedback regulation is fairly
abstract and general enough to have many alternative biochemical
implementations. It is possible, for example, for the enzyme activity
to be allosterically enhanced by the cooperative binding of $C$ or $P$
(termed substrate and product activation respectively in the language
of enzyme kinetics), giving rise to a Hill-like relation between
effector abundance and enzyme activity \cite{Sauro2012}.  In this work
we will work with the abstract activation rate function defined above.

In the following we analyze the closed-loop SFE network behavior by
studying how the SFE network properties described in the previous
sections are transformed under feedback.

\subsubsection{Feedback results in a dramatic noise reduction and
  increased robustness to parameter variation}

Here we study the SFE network under the influence of negative
feedback. We should point out that we characterize the
  open- and closed-loop systems with respect to the same features
  (noise and robustness), not to directly compare them, but because
  these features play an important role in the function of both
  mechanisms. Whenever we use the open-loop system as a baseline for
  assessing closed-loop system properties, scale-independent measures
  are used since this allows for the evaluation of relative
  distribution spreads. This principle is only disregarded in
  Fig.~\ref{N_pert_Cfeed} below.

For our first test, we use the same settings and parameters as those
of Fig.~\ref{2dscan}(a), only this time we add a feedback term from the
substrate to the enzyme activation rate.
Increasing the gain $K$ or shifting the activation point $x_0$ to the
left results in a decrease of both means and variances of substrate
and product. For the ranges of $K$ and $x_0$ values
  considered, the means change by at most a factor of 2.5, while the
  variances by about 5 times. At the same time, the CVs vary by about
  50\% and the corresponding Fano Factors by a factor of 5. Moreover,
as the analysis in the Supplement (section 5) shows, the CVs of both species
become relatively flat as $K$ and $x_0$ increase, while the Fano
Factor gets very close to 1 as $K$ increases for small values of
$x_0$, indicating that the resulting substrate and product stationary
distributions are approximately Poissonian in this regime. For the
analysis that follows, we fix $K=3$ and $x_0=5$ in the feedback
function $f(\cdot)$.

With the above choice of feedback parameters, we first study the
sensitivity of the closed-loop SFE system to variations of the two key
parameters, $\alpha_E$ and $k$.  As Fig.~\ref{2dscan}(b)
demonstrates, means and variances (and, consequently, CVs) of
substrate and product become largely independent of $\alpha_E$, except
for very large values of $k$ (similar results are obtained for product
feedback). Moreover, noise of substrate and product is dramatically
reduced in comparison to the open-loop SFE system, while the variation
of means and variances is now quite small, despite the large ranges of
$\alpha_E$ and $k$ values considered. It is also worth noting that the
Fano Factors of both substrate and product are very close to one for a
large range of parameters.


Interestingly, if we quantify noise reduction by the ratio of
open-loop vs.~closed-loop CV, we observe that noise reduction is
maximal where the open-loop SFE system noise is greatest, as demonstrated
by comparing Fig.~\ref{closed_scheme}(c) with Fig.~\ref{2dscan}(a).


Another striking effect of feedback regulation of enzyme activity is
that the closed-loop SFE system becomes much less sensitive to
parameter variations in comparison to the open-loop case. Applying the
same parametric perturbations described in \S\ref{sub_pert_open}, we
obtain the histograms of Fig.~\ref{fig:pert_open_closed} (continuous lines). As it
becomes apparent, the histograms corresponding to the closed-loop SFE
system are several orders of magnitude narrower compared to the
open-loop. Moreover, despite the relatively large parametric
perturbations, variability in substrate and product statistics of the
closed-loop SFE system is largely contained within an order of
magnitude.


As it was pointed out in \S\ref{sub_pert_open}, variability in
biochemical reaction rates can be considered ``artificial'', however
changes in the number of enzymes, $N$, are to be expected in a
cellular population. It is therefore interesting to study how
variations in $N$ alone are propagated to the substrate and product
statistics. Assuming that both the open- and closed-loop systems
operate with the same average number of active enzymes for the
``nominal'' value of $N=8$, Fig.~\ref{N_pert_Cfeed}(a,b) shows how the
substrate mean and variance vary as $N$ is perturbed around this
value, both in the open- and closed-loop systems (with substrate
feedback). To achieve the same average number of active enzymes for
$N=8$, the closed-loop SFE system was simulated first, and the mean
number of active enzymes was recorded. This number was then used to
back-calculate an appropriate $\alpha_E$ value (keeping $\mu_E$ fixed)
for the open-loop SFE system. Panel (c) also shows how the
distribution of active enzymes differs in the two systems for
$N=8$. The cyan line corresponds to a binomial distribution,
$\mbox{B}(8,p)$, where $p = \alpha_E/(\alpha_E+\mu_E)$ is determined
by the $\alpha_E$ and $\mu_E$ values of the open loop. The red
distribution is obtained from simulation of the closed loop and is
markedly different from a binomial. The difference is especially
significant at the lower end, as small values of $E^*$ lead to fast
accumulation of $C$. Similar results are obtained for product
feedback.

\begin{figure}[h!tb]
\centering
  \includegraphics[width=0.8\textwidth]{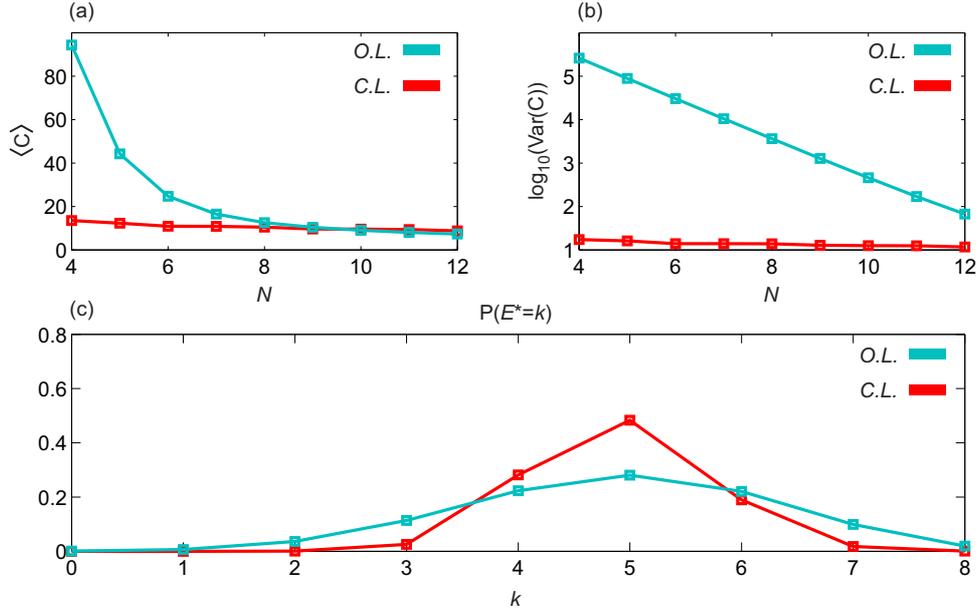}
  \caption{\textbf{(a, b)} Mean and variance of substrate in the open-
    and closed-loop SFE system with substrate feedback (notice the
    logarithmic variance scale). The closed loop was simulated with
    $\alpha_E=0.04$, $\mu_E=0.1$, $\alpha_C=5010$, $\nu=1$, $k=100$,
    $\mu_P=1$ and feedback parameters $K=10$, $C_0=15$.  In the
    open-loop SFE system all parameters were kept the same, except for
    $\alpha_E$, which was set to 0.1572 to achieve the same mean of
    $E^*$ that the closed-loop SFE system achieves for $N=8$.}
  \label{N_pert_Cfeed}
\end{figure}

\subsubsection{Open-loop stochasticity vs.~closed-loop determinism}

A further remarkable by-product of feedback in the closed-loop SFE system
is the fact that the mean of the stochastic model ends up following
very closely the predictions of the ODE equations for the
deterministic system. This behavior becomes more pronounced as the
number of available enzymes ($N$) grows, while the average number of
active enzymes ($N\alpha_E/(\alpha_E+\mu_E)$) remains small. Under
this condition, one can think of enzyme activation in the original
system as a zeroth-order reaction with rate $N\alpha_E$, and the
active enzyme abundance to be described by a birth-death process with
birth rate $N\alpha_E$, death rate $\mu_E$ and Poisson stationary
distribution with parameter $\lambda=N\alpha_E/\mu_E$.

The accuracy of the ODE approximation to the mean substrate levels in
the case of substrate feedback can be demonstrated using the same type
of parametric perturbations with those employed in
\S\ref{sub_pert_open}. All nominal parameters were kept the same for
this test, except for $\mu_E$, $N\alpha_E$ and $k$, which were set
equal to 1, 5 and 100 respectively ($N=100$, $\alpha_E=0.05$). We
compare the mean of the stochastic model with the ODE prediction in
the case of substrate feedback with $K=0.4,~x_0=5$, and define the
relative error
\begin{align}
  R.E. &:= \frac{|\langle C\rangle-C_{ss}|}{\langle C\rangle}\cdot 100\%,
\end{align}
where $C_{ss}$ is the equilibrium solution of the ODE model. In this
setting, a set of 5000 random perturbations leads to an average
relative error of 1.5\% with standard deviation 0.96\%, which clearly
shows that the ODE solution captures the mean substrate abundance with
very good accuracy indeed (note that the same holds for the mean of
$P$, since it depends linearly on the mean of $C$). Very similar
results are obtained in the case of product feedback.

The above observations are even more striking, if we take into account
1) the fact that the closed-loop SFE system is still highly nonlinear
and 2) the intrinsic property of stochastically focused systems to
display completely different mean dynamics when compared to the ODE
solutions. An explanation of this behavior can be given by examining
the moment differential equations. In the limiting case considered in
this section, denoting $\alpha_E(C)$ the production rate of active
enzyme, we obtain
\begin{align}
  \frac{d\br{C}}{dt}&= \alpha_C-\nu\br{C}-k\br{E^*C}=\alpha_C-\nu\br{C}-k\mbox{Cov}(E^*,C)-k\br{C}\br{E^*}\\
  \frac{d\mbox{Cov}(E^*,C)}{dt} &=
  -(\mu_E+\nu)\mbox{Cov}(E^*,C)-\mbox{Cov}(kE^*C,E^*)+\mbox{Cov}(\alpha_E(C),C)\label{covEC}
\end{align}

The last two terms on the right-hand side of \eqref{covEC} denote the
covariance of the substrate enzymatic degradation rate with active
enzyme and the covariance of the enzyme activation rate with the
substrate. Both covariances are expected to be positive at steady
state, which implies that the terms act against each other in
determining the steady-state covariance of substrate and active
enzyme. In turn, a small value of this covariance (compared to the
product $\br{C}\br{E^*}$) implies that the mean of $C$ can be
approximately captured by a mean-field equation, where $\br{E^*C}$ has
been replaced by $\br{C}\br{E^*}$. This is indeed the case in our
simulations, where $\mbox{Cov}(E^*,C)$ turns out to be $\sim$20--30
times smaller than $\br{C}\br{E^*}$. On the other hand, the open-loop
value of $\mbox{Cov}(E^*,C)$ is about 30 times larger than the
closed-loop one. This comes as no surprise, as one expects substrate
and active enzyme to display a strong negative correlation, which is
the cause of the discrepancy between stochastic and deterministic
descriptions of stochastically focused systems.

Similar observations can be made when $N$ is small (e.g.~around 10),
however the relative errors become at least one order of magnitude
larger. We believe that this can be attributed to the fact that the
enzyme activation propensity depends both on the abundance of inactive
enzyme and the substrate/product abundance, which increases
the inaccuracy of the ODEs.

\subsubsection{The feedback mechanism is intrinsically robust}

\begin{figure}[h!tb]
\centering
  \includegraphics[width=0.8\textwidth]{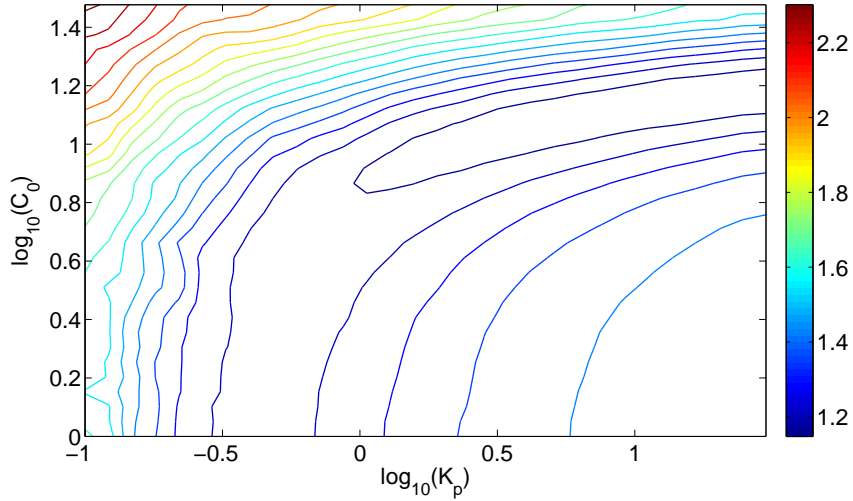}
  \caption{Contour lines of $\log_{10}(h(K,C_0))$ for $C_{target}=10$,
    obtained by evaluating the function via stochastic simulation on a
    logarithmic grid in the parameter space.}
  \label{optimal_feedback}
\end{figure}

As we have already demonstrated, the closed-loop SFE system is
remarkably robust to parametric perturbations of the open-loop
model. However, in all of our numerical experiments we have kept the
parameters of the feedback function $f$ fixed to a few different
values. Here we examine the opposite situation, in which only the
controller parameters are free to vary while the rest are held
constant. We therefore consider the problem of regulating the mean of
$C$ around a fixed value with feedback from $C$. The problem can be
posed as follows:
\[ \underset{K,~C_0}\min~~ h(K,~C_0):= \br{(C-C_{target})^{2}} = (\br{C}_{ss}-C_{target})^2+\mbox{Var}(C)_{ss},\]
where both the mean and variance of $C$ depend on the feedback
function parameters. Fig.~\ref{optimal_feedback} shows the contour
lines of $h$, obtained via stochastic simulation over a wide range of
$K$ and $C_0$ values for $C_{target}=10$. It can be observed that $h$
is more sensitive to $C_0$ than $K$: beyond a certain $K$ value, the
function quickly levels off.

Based on our simulation runs, the optimal feedback parameters turned
out to be $K=30$ (the maximum $K$ value considered for the plot) and
$C_0\simeq 16$ (given the inevitable uncertainty in $h$ due to
sampling variability, the true optimal value should be close to
this). The optimal feedback function therefore resembles a
``barrier'': for $x<C_0$ it is zero, while it rises very steeply
beyond $C_0$.

Note that the mean and variance of $C$ both depend on $(K,C_0)$, but
neither quantity is available in closed form as a function of the
feedback parameters or obtainable from a closed set of moment
equations. Thus, $h$ had to be evaluated on a grid with the help of
stochastic simulation. Alternatively, as we show in the Supplement (section 6),
one can exploit the behavior presented in the previous Section, and
optimize a similar objective function by directly evaluating the
required moments of $C$ using a simple moment closure
approximation based on the \emph{method of moments}. This scheme,
introduced in \cite{Engblom06}, provides very accurate approximations
of the mean and variance at a fraction of the computational effort,
thus allowing optimization to be carried out very efficiently. The
optimal parameters for the approximate system can be used as starting
points in the optimization of $h$.


\section{Discussion}

In this work we have examined the behavior of a branched enzymatic
reaction scheme. This system has already been shown to display
stochastic focusing, a sensitivity amplification phenomenon that
arises due to nonlinearities and stochasticity whenever only a few
enzyme molecules are present in the system. We have additionally shown
that when the enzyme activity evolves on a slow timescale compared to
that of the substrate, very large fluctuations can be generated in the
system. Moreover, the dynamics of the system is extremely sensitive to
variations in its reaction rates. Both these observations imply that
this simple model is not appropriate for robust signal detection.

We asked how the system behavior would change in the presence of a
feedback mechanism, so that the ``controller'' molecule ($E^*$) could
sense the fluctuations in $C$ (its substrate) or $P$ (the product of
the alternative reaction branch). We have shown that noise decreases
dramatically in the presence of feedback, while the robustness of the
average system behavior is boosted significantly. Consequently, the
focused system with feedback ends up behaving almost as predictably as
a mean-field ODE model, even when the number of active enzymes is very
small.

There exist several biochemical systems which in certain
  aspects match the main ideas behind the SFE motif, i.e.~display
  stochastic fluctuations in enzyme activity/abundance and
  substrate/product feedback activation. For example, it was recently
  discovered \cite{Iversen14} that the guanine nucleotide exchange
  factor SOS and its substrate, the Ras GTPase, are involved in a
  feedback loop where SOS converts Ras-GDP (inactive) to Ras-GTP
  (active) and, in turn, active Ras allosterically stabilizes the
  high-activity state of SOS. Another prominent example is microRNA
  post-transcriptional regulation of gene expression \cite{Tsang07},
  where a microRNA may mediate the degradation of a target mRNA, while
  the protein arising from this mRNA in turn activates the microRNA
  transcription.

Yet another instance is the heat shock response system in
\emph{E. coli} \cite{ElSamad05}: here the $\sigma^{32}$ factor, which
activates the heat-shock responsive genes, is quickly turned over
under the action of the protease FtsH at normal growth
temperatures. After a shift to high temperature, $\sigma^{32}$ is
rapidly stabilized
and at the same time it also activates the synthesis of FtsH. The
FtsH-mediated degradation of $\sigma^{32}$ is under negative
feedback. Finally, it is known that mRNA decapping (a process that
triggers mRNA degradation) is controlled by the decapping enzyme Dcp2,
which fluctuates between an open (inactive) and closed (active)
form. Experimental evidence suggests that the closed conformation of
the enzyme is promoted by the activator protein Dcp1 together with the
mRNA substrate itself \cite{Floor08,Deshmukh08}. We should stress,
however, that it is still unclear if any of the aforementioned
examples display all the dynamic features of the motif considered in
this work. Speaking generally, due to the required levels of
measurement accuracy, it is difficult to find examples that exactly
match the conditions considered here with current experimental
techniques. However, it is certainly conceivable that this will be
achieved in the future.

An interesting feature of our system is that homeostasis is achieved
with a very small number of controller molecules (in the order of 10),
which are able to maintain the output at a very low level with
fluctuations that are -- to a good approximation -- Poissonian. We
have considered two alternative feedback schemes: in the first the
substrate directly affects the activation rate of the enzyme (a case
of substrate activation), while in the second the product of the
alternative reaction branch is used as a ``proxy'' for the substrate
abundance. Note that the flux through the $C$-$P$ branch is many
orders of magnitude smaller than the flux in the $E$-regulated
branch. The $C$-$P$ branch can thus be thought to act as a ``sensor
mechanism'', used to control the high-flux branch of the system. To
the best of our knowledge, this type of feedback has not yet been
observed in naturally occurring reactions.

Finally, it is worth to note that one could achieve this type of
regulation with an ``unfocused'' system, in which the coupling between
enzyme and substrate (parameter $k$) would be much smaller. This would
imply, however, that the number of enzyme molecules needed to achieve
the same substrate levels would have to be much greater, and this
could entail an added cost for a living cell.

In summary, to regulate a low-copy, high-flux substrate via an
enzymatic mechanism such as the one considered here, there are three
possibilities: a) use of a low number of controller enzymes and strong
coupling between enzyme and substrate (which results in stochastic
focusing and noise), b) use of a high-copy enzyme and weak coupling
(with the associated production cost) or c) use of a low-copy
controller with feedback: an alternative which, as we have
demonstrated, leads to a remarkably well-behaved closed-loop system.

It is thus conceivable that cellular feedback mechanisms have evolved
to exploit the nature of stochastically focused systems to achieve
regulation of low-copy substrates with the minimal number of
controller molecules. We expect that the rapid development of
experimental techniques in single-molecule enzymatics will soon enable
the experimental verification of our findings and possibly the
discovery of similar noise attenuation mechanisms inside
cells. Finally, our results can be seen as a first step towards the
rational manipulation of noise properties in low-copy enzymatic
reactions.


\section*{Acknowledgments}

This work was financially supported in part by the Swiss National Science Foundation (A.M.-A. and M.K.)
and the Swedish
Research Council within the UPMARC Linnaeus center of Excellence
(S.E. and P.B.)


\newcommand{\doi}[1]{\href{http://dx.doi.org/#1}{doi:#1}}
 \newcommand{\noop}[1]{}

\newcommand\supplcite[1]{\mbox{\citep[S][]{#1}}}

\newpage

\appendix
\begin{center}
  \LARGE{\textbf{Stochastic focusing coupled with negative feedback enables robust regulation in biochemical reaction networks:\\Supplementary Material}}
\end{center}

\section{Some comments on the choice of the reaction scheme}
\subsection{Theoretical analysis}
The main point of our analysis here is to determine the \emph{sensitivity} of a branched-reaction product to changes in the activation rate of an enzyme and in this way provide some justification for our modeling choices. The rationale behind this analysis is that one cannot hope to control the mean - let alone the variance - of a product, if its statistics are not sensitive to changes in the enzyme. With this in mind, we examine the following branched reaction system:

\begin{align*}
&\varnothing \rates{\alpha_E}{\mu_E} E\notag \\
&\varnothing \xrightarrow{\alpha_C} C \\
& C \xrightarrow{\nu }P\xrightarrow{\mu_{P}}\varnothing\\
&C+E \xrightarrow{k} Q +E\notag\\
& Q\xrightarrow{\mu_{Q}}\varnothing\\
\end{align*}

The system consists of the following:
\begin{itemize}
\item An enzyme $E$, that is found in low copy numbers and therefore its fluctuations have a significant impact on system behavior.
\item A high-copy, low-noise enzyme (not shown), responsible for the conversion of $C$ to $P$. Alternatively, we can assume that $C$ ``matures'' into $P$ without the help of an enzyme. In both cases, this reaction can be considered to be first-order, even if it is enzymatic.
\item Two products, $P$ and $Q$, which are produced from $C$
\item The substrate species, $C$, which plays the most critical role. $C$ enters the system through a zeroth-order reaction, and can have two alternate fates: it can either be converted to $P$ or $Q$.
\end{itemize}

The initial sensitivity question can be now posed more precisely: which of the two reaction products, $P$ or $Q$, is more sensitive to changes in the activation rate of $E$? Apart from the system structure, we are making the following assumptions regarding the reaction rates:

\begin{itemize}
\item The bimolecular reaction rate ($k$) is large compared to the
  first-order reaction rate of $C$ ($\nu$). That is, most of the
  influx of $C$ is directed towards $Q$. This assumption amplifies the effect of the
  nonlinear kinetics in the system (in the opposite case, the bimolecular reaction could
  be considered as a perturbation in an almost-linear network).

\item The influx of $C$ to the system is high, i.e. $\alpha_C$ is also
  large.

\item The rates of $E$ are such that $E$ has low copies and high noise
  (this was already stated above), so that we cannot replace $E$ by
  its mean in the bimolecular reaction.
\end{itemize}

We now want to see what happens to the steady-state means $\langle
P\rangle_{ss}$ and $\langle Q\rangle_{ss}$ when $\alpha_E$ varies. In
the case of $\langle P\rangle_{ss}$, the situation is simple: $\langle
P\rangle_{ss}$ follows the behavior of $\langle C\rangle_{ss}$,
\[\langle P\rangle_{ss}=\frac{\nu}{\mu_{P}}\langle C\rangle_{ss}.\]

Thus, our focus shifts from $P$ to $C$ in this case. If the mean of $C$ is sensitive to changes in $E$, we know that $P$ will be sensitive as well. This is precisely the case when stochastic focusing is present.

In the case of $Q$, the situation is different: we expect it to be so, because in order to produce a $Q$ molecule, we need both $E$ and $C$ to be present. Therefore, if $E$ hits zero, $C$ will inevitably accumulate (since we assumed that the rate of the alternate path, $\nu$, is small), but no $Q$ will be produced. Instead, while $E$ stays at zero, $Q$ will \emph{drop} with a speed that depends on $\mu_{Q}$. Once $E$ returns to non-zero numbers, the accumulated amount of $C$ will be converted into $Q$ in a strong production burst. Depending on $\mu_{Q}$, this may result in a brief burst of $Q$, or may go unnoticed (when $\mu_{Q}$ is small enough, the burst will happen, because $Q$ cannot be removed fast enough from the system). We thus see that $Q$ can display a more complex behavior than $P$.

We next turn to the mean of $Q$. Since $\nu\ll k$, let us assume first that $\nu\simeq 0$. In this case, the first-moment equation for $C$ will give
\[ \langle EC\rangle_{ss}=\frac{\alpha_C}{k}.\]

Note that the mean of $C$ does not appear there, because we assumed no first-order degradation. The equation says that the steady-state mean of the product of $E$ with $C$ is constant, \emph{independently of} $\alpha_E$. Turning to the equation of the first moment of $Q$, we then get
\[ \langle Q\rangle_{ss}=k\frac{\langle EC\rangle_{ss}}{\mu_{Q}}=\frac{k\alpha_C}{k\mu_{Q}}=\frac{\alpha_C}{\mu_{Q}},\]
which shows that $Q$ is also not affected in its mean by changes in
the rates of $E$. Intuitively, we can see why it is plausible for the
mean of $Q$ to remain constant, by looking at the bimolecular reaction
that produces it: when $\alpha_E$ increases, the mean of $C$ drops and
vice versa. Thus, the average production rate of $Q$ cannot change
that much --- and in fact, does not change at all in the limiting case
$\nu= 0$.

Using this observation, we can understand also why $\langle Q\rangle_{ss}$ is not sensitive to $\alpha_E$ when $\nu$ is non-zero but small: while in this case the above relations do not hold exactly, we still expect them to hold with good precision (simulation verifies that). Overall then, we see that $Q$ is relatively insensitive to changes in $\alpha_E$, compared to $P$, and it makes sense to consider $Q$ as the target for regulation.

It should be noted that the above arguments hold only for the \emph{means} of $P$ and $Q$. We do not expect the variance of $Q$ to be equally insensitive to the noise in $E$, however it is not entirely clear how $Q$ could be used in a noise-suppressing feedback mechanism.

\subsection{Numerical verification}

\subsubsection{The sensitivity of $Q$}

To get a feeling for the scaling of the different constants we
consider the equilibrium solutions to the ODE model of the
network. The following three relations are immediate:
\begin{align}
  \label{eq:c1}
  C_{ss} &= \frac{\alpha_C}{\nu+k E_{ss}},  \\
  \label{eq:e1}
  E_{ss} &= \frac{\alpha_E}{\mu_E},  \\
  \label{eq:p1}
  Q_{ss} &= \frac{k C_{ss} E_{ss}}{\mu_Q}.
\end{align}

Assuming the mean lifetime of the substrate and the enzyme to be
about the same we may pick the units of time such that $\mu_C = \mu_E
= 1$. As we are interested in stochastic focusing, a low copy-number
phenomenon, we further prescribe as a \emph{base case} that $C_{ss} = E_{ss}
= Q_{ss} = 10$. Combined with \eqref{eq:c1} and \eqref{eq:p1} this
implies
\begin{align}
  \label{eq:ac}
  \alpha_C &= 10 \cdot (1+10k), \\
  \label{eq:mp}
  \mu_Q &= 10k.
\end{align}
With the enzymatic rate parameter $k$ still free and \eqref{eq:ac}--\eqref{eq:mp} given, we can next consider the
rate for the inflow of enzyme $\alpha_{E}$ to be an adjustable
parameter which controls the amount of product $Q$.

Using \eqref{eq:c1}--\eqref{eq:p1} and \eqref{eq:ac}--\eqref{eq:mp} we
arrive at the relation
\begin{align}
  \label{eq:p3}
  Q_{ss} &= \alpha_E \frac{1+10k}{1+\alpha_E k}.
\end{align}
For any given rate $k$ and desirable setpoint $Q_{set}$, \eqref{eq:p3}
can be solved for the value of $\alpha_{E}$ that makes $Q_{ss} =
Q_{set}$.

We now define the \emph{gain} $g$ as the response to a 50\% decrease
of enzyme from the base case $\alpha_E = 10$,
\begin{align}
  \label{eq:gain1}
  g = Q_{ss}(\alpha_E=10)/Q_{ss}(\alpha_E=5)-1.
\end{align}
With $g \le 0$, $Q$ does not respond (i.e.~is insensitive), and $g = 1
= 10/5-1$ can be considered a perfect transmission.

In the table below values from \eqref{eq:gain1} using \eqref{eq:p3}
have been computed for different values of the rate constant $k$. The
conclusion is that $k \lesssim 1$ is required for the network to be
responsive.
\begin{center}
\begin{tabular}{ l  r r r r r}
  $k$              & $10^{-2}$ & $10^{-1}$ & $10^{0}$ & $10^{1}$ & $10^{2}$  \\
  \hline
  $Q_{ss}(\alpha_E=10)$ & 10    &  10    & 10    &  10     & 10 \\
  $Q_{ss}(\alpha_E=5)$   & 5.24 &  6.67 & 9.17 &  9.90  &  9.99 \\
  \hline
  $g$ & 0.9091 & 0.5000 & 0.0909 & 0.0099 & 0.0010
\end{tabular}
\end{center}

In the stochastic setting, we note that focusing occurs due to a large
rate constant $k \gg 1$ since this is the only nonlinearity present in
the model. The following numerical results were obtained after
averaging over 5000 trajectories.
\begin{center}
\begin{tabular}{ l  r r r r r}
  $k$              & $10^{-2}$ & $10^{-1}$ & $10^{0}$ & $10^{1}$ & $10^{2}$  \\
  \hline
  $\langle Q \rangle_{\alpha_E=10}$ & $10.01 \pm 3.04$ &  $9.85   \pm 3.40  $ &  $9.91   \pm 3.25  $ &  $9.99   \pm 3.18  $ &  $10.00 \pm 3.17  $ \\
  $\langle Q \rangle_{\alpha_E=5}$ & $7.13  \pm 2.48$ &  $6.53   \pm 2.967  $ &  $8.91   \pm 3.44  $ &  $9.84   \pm 3.44  $ &  $9.97    \pm 4.02  $ \\
  \hline
  $g$ & 0.4039 &  0.5087  &  0.1113  &  0.0152 &  0.0028
\end{tabular}
\end{center}

The conclusion is that neither the deterministic nor the stochastic
model of $Q$ is sensitive to changes in the enzyme when focusing is
present.

\subsubsection{The sensitivity of $P$}

As before we pick units of time such that $\mu_E=\nu=1$. We get from
the ODE equilibrium solutions that
\begin{align}
  \label{eq:c4}
  C_{ss} &= \frac{\alpha_C}{\nu +k E_{ss}}  \\
  E_{ss} &= \frac{\alpha_E}{\mu_E}  \\
  P_{ss} &= \frac{\nu C_{ss}}{\mu_P},
\end{align}
such that via the base case $C_{ss} = E_{ss} = P_{ss} = 10$ we arrive
at
\begin{align}
  \label{eq:c3}
  \alpha_C &= 10 \cdot (\nu+10k), \\
  \label{eq:p4}
  \mu_P &= \nu.
\end{align}

We now define the gain $g$ as
\begin{align}
  \label{eq:gain2}
  g = \langle P \rangle_{\alpha_E=5} / \langle P
    \rangle_{\alpha_E=10}-1
\end{align}
since the enzyme $E$ this time acts as an inhibitor. In the table
below we compute the gain for various values of $\nu$ with $k =
10^{2}$. It can be seen that, in the deterministic case, this
network has good transmission ($g \sim 1$) when $\nu \lesssim k$.

\begin{center}
\begin{tabular}{ l  r r r r r}
  $k$		& $10^{2}$  & $10^{2}$ & $10^{2}$ & $10^{2}$ & $10^{2}$  \\
  $\nu$		& $10^{-1}$ & $10^{0}$ & $10^{1}$ & $10^{2}$ & $10^{3}$	 \\
  \hline
  $g$ & 1 & 1 & 0.98 & 0.83 & 0.33
\end{tabular}
\end{center}

In the stochastic regime we performed many simulations for various
combinations of parameters $k$ and $\nu$. The table below summarizes
the most interesting results found in this way. The stochastic
focusing effect is clearly present in the observed increase of gain
compared to the deterministic model.

\begin{center}
\begin{tabular}{ l  r r r r r}
  $k$		& $10^{2}$  & $10^{2}$ & $10^{2}$ & $10^{2}$ \\
  $\nu$		& $10^{-1}$ & $10^{0}$ & $10^{1}$ & $10^{2}$ \\
  \hline
  $\langle P \rangle_{\alpha_E=10}$ & $10.65 \pm 3.39$ &  $11.32   \pm 5.31  $ &  $11.31   \pm 3.25  $ &  $11.01   \pm 5.18  $ \\
  $\langle P \rangle_{\alpha_E=5}$ & $22.41  \pm 32.49$ & $36.08   \pm 92.08  $ & $29.42   \pm 53.49  $ &  $21.80   \pm 12.61  $ \\
  \hline
  $g$ & 1.2410 &  2.6080  &  1.9420  &  1.1800 \\
\end{tabular}
\end{center}

\section{Considering the effect of enzyme saturation in the catalytic degradation reaction}

We here consider a more realistic enzymatic reaction mechanism for the
degradation of the substrate $C$, which includes the formation of an
enzyme-substrate complex. As we will see, this more detailed
mechanism implies an overall behavior which is similar to the SFE model studied in the main paper

\begin{figure}[h!tb]
  \begin{center}
    \includegraphics[width=0.5\columnwidth]{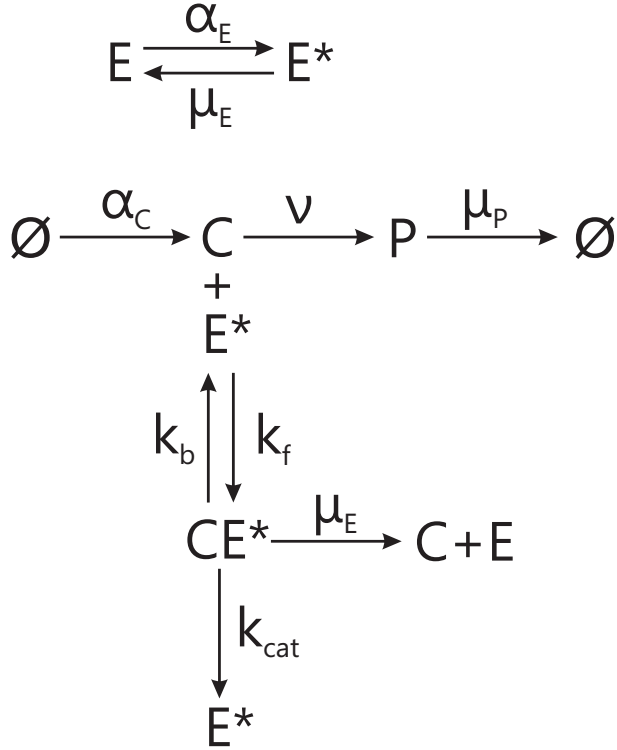}
  \end{center}
  \caption{A more detailed reaction mechanism for substrate degradation.}
  \label{fig:enzyme_saturation_mech}
\end{figure}

The mechanism is displayed on
Fig. \ref{fig:enzyme_saturation_mech}. According to this scheme,
active enzyme ($E*$) and substrate have to first form a complex before
$C$ is degraded. If the enzyme spontaneously switches to its inactive
form while bound to $C$, we assume that the complex dissociates.

With the additional reactions, the open-loop system becomes again
analytically intractable, as the conditional moment equations are no
longer closed. We will therefore base our analysis on the behavior of
the corresponding deterministic system. To further simplify the task,
we will first assume a fixed number of active enzymes,
$E^*_{tot}=E^*+CE^*$, i.e. ignore the enzyme (de)activation
reactions. Moreover, we will assume that the substrate flux towards
$P$ is much smaller than the flux into the enzymatic reaction and
therefore set $\nu=0$. Under these conditions, one can verify that the
steady-state concentration of substrate and free (active) enzyme are
given by the following expressions:

\begin{align}
C_{ss} &= \frac{\alpha_C(k_b+k_{cat})}{k_f(k_{cat}E^*_{tot}-\alpha_C)}\\
E^*_{ss} &= \frac{k_{cat}E^*_{tot}-\alpha_C}{k_{cat}}
\end{align}

We see that the existence of a finite positive steady-state for the
system depends on the relation of $E^*_{tot}$ with the ratio
$\alpha_C/k_{cat}$, which connects the rates of substrate influx and
the catalytic efficiency of the active enzyme. In other words, if
there is not a sufficient number of active enzymes in the system,
there is no finite and positive steady-state; instead, the number of
substrate molecules tends to infinity, as the existing free enzyme
molecules are completely saturated and cannot process all incoming
substrate molecules.

When an alternative fate is available for the substrate (e.g. its
conversion to $P$), the system will of course be stable, as the
alternative pathway will absorb the excess substrate. However, if
$\nu$ is small compared to $k_{cat}$, the resulting steady-state
substrate concentration, expected to be close to $\alpha_C/\nu$, will
be large.

Now, let us consider the case where $E^*_{tot}$ is varied
externally. As it approaches $\alpha_C/k_{cat}$ from above, the
steady-state concentration of $C$ quickly diverges to
infinity. Therefore, it is reasonable to imagine that if active
enzymes fluctuate slowly, randomly and close to the critical value
$\alpha_C/k_{cat}$, this will result in dramatic fluctuations in $C$.

\begin{figure}[h!tb]
  \begin{center}
    \includegraphics[width=0.7\columnwidth]{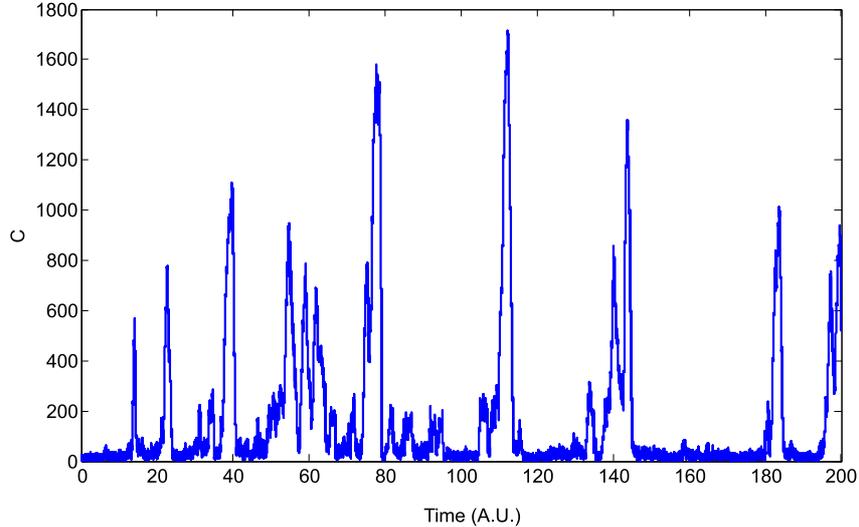}
  \end{center}
  \caption{A sample path of $C$ for the system of
    Fig. \ref{fig:enzyme_saturation_mech} for $\alpha_E=0.16$,
    $\mu_E=0.1$, $\alpha_C=5000$, $\nu=1$, $\mu_P=1$, $k_f=100$,
    $k_b=10$, $k_{cat}=1000$ and $N=10$ ($N$ is the total number of
    enzymes, as defined previously). With the selected rates the
    average number of active enzymes is close to 6, while
    $\alpha_C/k_{cat}=5$. As expected, we observe an amplification of
    the enzyme fluctuations in the substrate abundance.}
  \label{fig:C_saturation}
\end{figure}

The intuition obtained from the above observations is confirmed by
stochastic simulation of the system, as shown on
Fig. \ref{fig:C_saturation}.

In summary, the open-loop SFE network with the more detailed enzymatic
reaction mechanism displays a behavior analogous to the simplified
system considered in the main paper, with the only difference that
when enzyme saturation is taken into account \emph{even a non-zero
number active enzymes may be insufficient to prevent large substrate
fluctuations}, if the enzyme is not much faster compared to the rate
of substrate influx (i.e. $k_{cat}\gg \alpha_C$).

Provided the total number of available enzymes is greater than the
necessary minimum to prevent complete saturation, we expect that
addition of feedback from $C$ or $P$ to the enzyme will, similarly to
the simplified case, result in great noise reduction. For the example
presented on Fig.~\ref{fig:C_saturation}, the CV of $C$ was reduced
from about 2 to about 0.38 and the Fano Factor from about 900 to about
1.73 for $K =50$ and $C_0=20$. Remarkably, the steady-state means of
the system were again very close to the ODE steady-state (obtained
from numerical simulation of the deterministic system): $\langle
C\rangle_{ss} \approx 11.5$, $\langle E^*\rangle_{ss} \approx 4.6$ and
$\langle E \rangle_{ss}= 0.4$, whereas $C_{ss} = 10.52$,
$E^*_{ss}=4.78$ and $E_{ss}=0.22$.


\section{Conditional moment equations for the open-loop system}

Denote $E^*$ by $Y$, $C$ by $Z_1$ and $P$ by $Z_2$. Since the total
number of enzymes, $E+E^*$ is assumed equal to $N$, we can write the
dynamics of $E^*$ without reference to $E$:
\[
\varnothing \xrightarrow{\alpha_E(N-E)}E\xrightarrow{\mu_E}\varnothing.
\]
We also know that the stationary distribution of $E^*$ is
$\mbox{B}(N,p)$, where $p=\frac{\alpha_E}{\alpha_E+\mu_E}$. We can
thus describe the evolution of moments of $C$ conditioned on the state
of $E^*$, following the approach described in \cite{Hasenauer13}. We
further simplify the problem by considering the steady-state
conditional moments. Following the notational conventions of
\cite{Hasenauer13}, we set
\begin{align*}
&\lim_{t\to\infty}\mathbb{E}[Z_1|y,t]:={\color{red}\mu_1(y)}\\
&\lim_{t\to\infty}\mathbb{E}[Z_2|y,t]:={\color{orange}\mu_2(y)}\\
&\lim_{t\to\infty}\mathbb{E}[\left(Z_1-{\color{red}\mu_1(y,t)}\right)^2|y,t]:={\color{blue}C_{(2,0)}(y)}\\
&\lim_{t\to\infty}\mathbb{E}[\left(Z_2-{\color{orange}\mu_2(y,t)}\right)^2|y,t]:={\color{magenta}C_{(0,2)}(y)}\\
&\lim_{t\to\infty}\mathbb{E}[\left(Z_1-{\color{red}\mu_1(y,t)}\right)\left(Z_2-{\color{orange}\mu_2(y,t)}\right)^2|y,t]:={\color{teal}C_{(1,1)}(y)}\\
&\lim_{t\to\infty}\mathbb{P}[Y(t)=y]:=p(y)
\end{align*}

The steady-state first- and second-order conditional moments of $C$
and $P$ are then obtained by solving the following system of linear
equations:

\begin{align}
&\mbox{\bf Steady-state equation for }{\color{red}\mu_1(y)}: \notag\\
&\alpha_Cp(y)-{\color{red}\mu_1(y)}p(y)\left(\alpha_E(N-y)+\mu_Ey+\nu+ky\right)+\alpha_E(N-y+1){\color{red}\mu_1(y-1)}p(y-1)+\notag\\
&+\mu_E(y+1){\color{red}\mu_1(y+1)}p(y+1)=0\\
&\mbox{\bf Steady-state equation for }{\color{orange}{\color{orange}\mu_2(y)}}: \notag\\
&\nu{\color{red}\mu_1(y)}(y)p(y)-{\color{orange}{\color{orange}\mu_2(y)}}p(y)\left(\alpha_E(N-y)+\mu_Ey+\mu_P\right)+\alpha_E(N-y+1){\color{orange}\mu_2(y-1)}p(y-1)+\notag\\
&\mu_E(y+1){\color{orange}\mu_2(y+1)}p(y+1)=0\\
&\mbox{\bf Steady-state equation for }{\color{blue}C_{(2,0)}(y)}: \notag\\
&\alpha_Cp(y)+\alpha_E(N-y+1)\left({\color{red}\mu_1(y-1)}-{\color{red}\mu_1(y)}\right)^2p(y-1)+\mu_E(y+1)\left({\color{red}\mu_1(y-1)}-{\color{red}\mu_1(y)}\right)^2p(y+1)+\notag\\
&+\nu{\color{red}\mu_1(y)}p(y)+ky{\color{red}\mu_1(y)}p(y)-{\color{blue}C_{(2,0)}(y)}p(y)\left(\alpha_E(N-y)+\mu_Ey+2ky+2\nu\right)+\notag\\
&+\mu_E(y+1){\color{blue}C_{(2,0)}(y+1)}p(y+1)+\alpha_E(N-y+1){\color{blue}C_{(2,0)}(y-1)}p(y-1)=0\\
&\mbox{\bf Steady-state equation for }{\color{magenta}C_{(0,2)}(y)}: \notag\\
&\alpha_E(N-y+1)\left({\color{orange}\mu_2(y-1)}-{\color{orange}\mu_2(y)}\right)^2+\mu_E(y+1)\left({\color{orange}\mu_2(y+1)}-{\color{orange}\mu_2(y)}\right)^2+\nu{\color{red}\mu_1(y)}p(y)+\notag\\
&+\mu_P{\color{orange}\mu_2(y)}p(y)-{\color{magenta}C_{(0,2)}(y)}p(y)\left(\alpha_E(N-y)+\mu_Ey+2\mu_P\right)+\mu_E(y+1){\color{magenta}C_{(0,2)}(y+1)}p(y+1)+\notag\\
&+\alpha_E(N-y+1){\color{magenta}C_{(0,2)}(y-1)}p(y-1)+2\nu {\color{teal}C_{(1,1)}(y)}p(y)=0\\
&\mbox{\bf Steady-state equation for }{\color{teal}C_{(1,1)}(y)}: \notag\\
&\alpha_E(N-y+1)\left({\color{red}\mu_1(y-1)}-{\color{red}\mu_1(y)}\right)\left({\color{red}\mu_2(y-1)}-{\color{red}\mu_2(y)}\right)p(y-1)+\notag\\
&+\mu_E(y+1)\left({\color{red}\mu_1(y-1)}-{\color{red}\mu_1(y)}\right)\left({\color{red}\mu_2(y-1)}-{\color{red}\mu_2(y)}\right)p(y+1)\mu_P{\color{red}\mu_2(y)}p(y)-\nu{\color{red}\mu_1(y)}p(y)-\notag\\
&-{\color{teal}C_{(1,1)}(y)}p(y)\left(\alpha_E(N-y)+\mu_Ey+\nu+ky+\mu_p\right)+\mu_E(y+1){\color{teal}C_{(1,1)}(y+1)}p(y+1)+\notag\\
&+\alpha_E(N-y+1){\color{teal}C_{(1,1)}(y-1)}p(y-1)+\nu {\color{blue}C_{(2,0)}(y)}p(y)=0
\end{align}

This system of equations has to be solved for all $y\in [0,N]$ to
yield $\mu_1(y)$, $\mu_2(y)$, $C_{(2,0)}(y)$, $C_{(0,2)}(y)$ and
$C_{(1,1)}(y)$, which in turn can be marginalized over $y$ to derive
unconditional moments. For example, \[\mu_1 :=
\lim_{t\to\infty}\mathbb{E}[Z_1(t)]=\sum_{y=0}^N \mu_1(y)p(y)\] and
\begin{align*}
C_{(2,0)}&:=\lim_{t\to\infty}\mathbb{E}[(Z_1(t)-\mu_1)^2]=\lim_{t\to\infty}\mathbb{E}[Z_1(t)^2]-\lim_{t\to\infty}\mathbb{E}[Z_1(t)]^2=\\
&=\sum_{y=0}^N\left(C_{(2,0)}(y)+\mu_1(y)^2\right)p(y)-\left(\sum_{y=0}^N\mu_1(y)p(y) \right)^2.
\end{align*}
In case the distribution of $y$ is not finitely supported (but still
known analytically), one can similarly solve over a finite set of $y$
values for $N$ large enough to capture the bulk of the probability
mass of $y$.

We should stress that the above system of linear equations is
\emph{exact}, i.e. no closure has been employed. As shown in
\cite{Hasenauer13}, the conditions for obtaining closed moment
equations are different in the conditional and unconditional
cases. The system studied here has non-closed unconditional moments, a
feature that has so far hindered the analytical study of stochastic
focusing. As can be seen, however, the conditional moment equations
are closed.


\clearpage

\section{Comparison of Mass Fluctuation Kinetics with the exact solution for the open-loop system}

\begin{figure}[!htb]
  \begin{center}
    \includegraphics[width=0.8\linewidth]{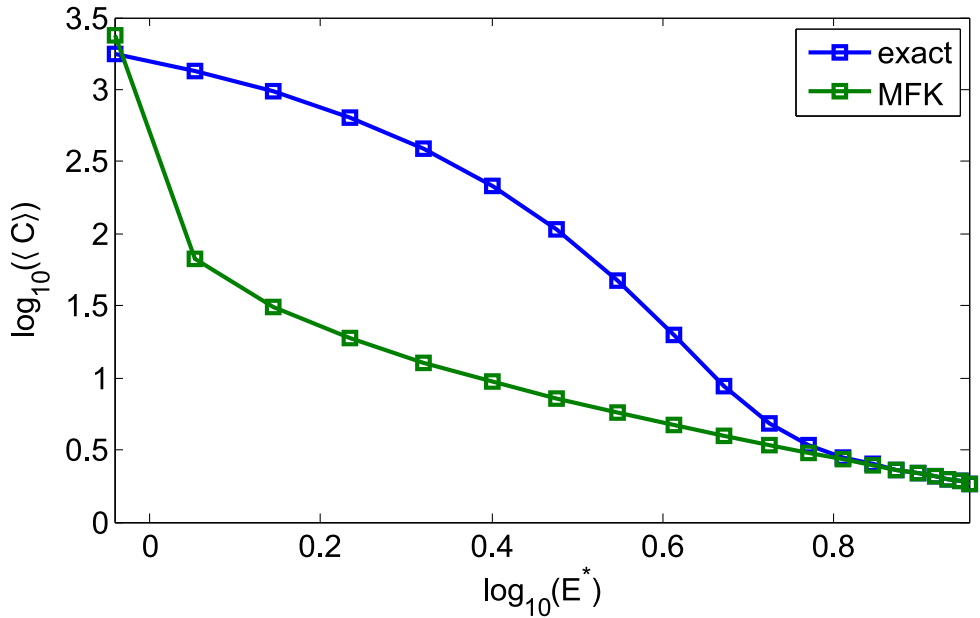}
  \end{center}
  \caption{Steady-state substrate mean as a function of the average number of active enzymes. The figure was obtained with the same parametrization used in Fig. 2 of the main paper. It can be clearly seen that in the region where stochastic focusing is observed, the MFK mean grossly underestimates the the true mean of the substrate. However, as the average number of active enzymes increases, MFK becomes increasingly exact.}
  \label{fig:MFK_means}
\end{figure}

\begin{figure}[!htb]
  \begin{center}
    \includegraphics[width=0.8\linewidth]{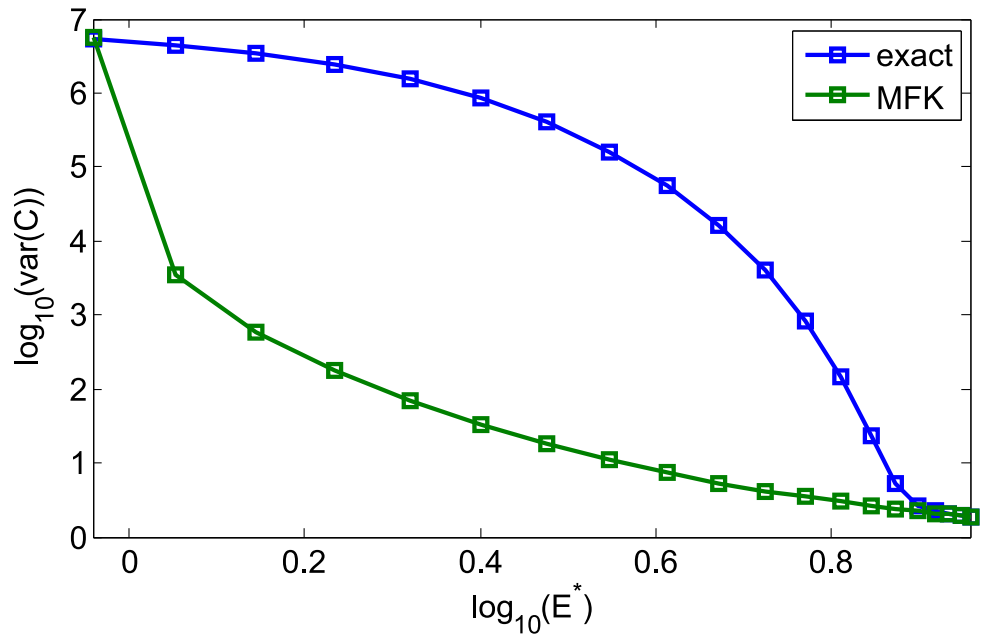}
  \end{center}
  \caption{Steady-state substrate variance as a function of the average number of active enzymes. The figure was obtained with the same parametrization used in Fig. 2 of the main paper. We observe that, besides underestimating the mean, MFK also underestimates the the true variance of the substrate. Again, as the average number of active enzymes increases, MFK becomes increasingly exact.}
  \label{fig:MKF_vars}
\end{figure}

Mass Fluctuation Kinetics \cite{MFK} is a popular moment closure technique that is used to approximate the evolution of means, variances and covariances in stochastic chemical kinetics. The approximate moment equations are derived by setting to zero the third-order cumulants (equal to the third order central moments) of all species. Below (Figs. \ref{fig:MFK_means} and \ref{fig:MKF_vars}) we show a comparison of the MFK approximation of mean and variance of $C$ with the exact solution based on conditional moments (open loop system). We can clearly see that MFK greatly underestimates both the mean and the variance of the substrate (notice the log scale on the y-axis), which proves that stochastic focusing cannot be adequately studied with this approximation. In fact, all moment closure methods tried on the system failed. Most likely, this happens due to the fact that in the presence of stochastic focusing the distributions of $C$ and $P$ become extremely skewed and consequently violate all commonly made regularity assumptions on which moment closure is typically based.

\clearpage
\section{Exploring the effect of feedback parameters on substrate and product statistics}

Here we present simulation results that explore how the behavior of the SFE network statistics changes under negative feedback. More specifically, we study the effect of feedback on the mean, variance CV and Fano Factor of substrate and product. This analysis helps to put in context our specific choice of feedback parameters used in the main text.

\begin{figure}[!ht]
  \begin{center}
    \includegraphics[width=\linewidth]{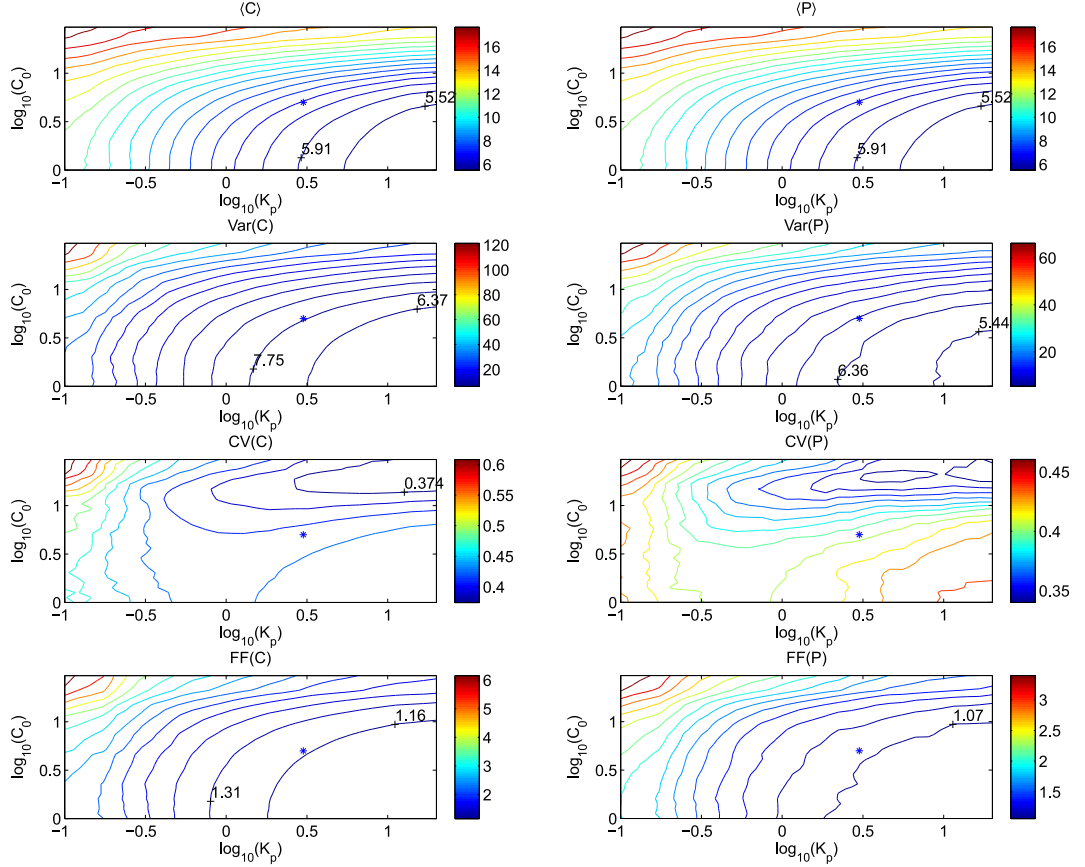}
  \end{center}
  \caption{Changes in $C$ and $P$ statistics generated by a scan over
    feedback parameters ($K_p$ and $C_0$) under feedback from $C$. As
    the strength of negative feedback increases (i.e. $K_p$ and $C_0$
    grow), both the mean and variance drop. However, the CV and Fano
    Factor behave differently: CV appears more sensitive to $C_0$ than
    $K_p$, while the Fano Factor depends equally on both
    parameters. Moreover, the two noise measures become minimal over
    different regions of the parameter space. As expected, the
    behavior of substrate fluctuations as parameters vary, is
    reflected in product fluctuations as well. The feedback parameter
    set used in the greatest part of the paper ($K_p=3$, $C_0=5$) is
    denoted by an asterisk.}
  \label{fig:feedback_scan_Cfeed}
\end{figure}

\begin{figure}[H]
  \begin{center}
    \includegraphics[width=\linewidth]{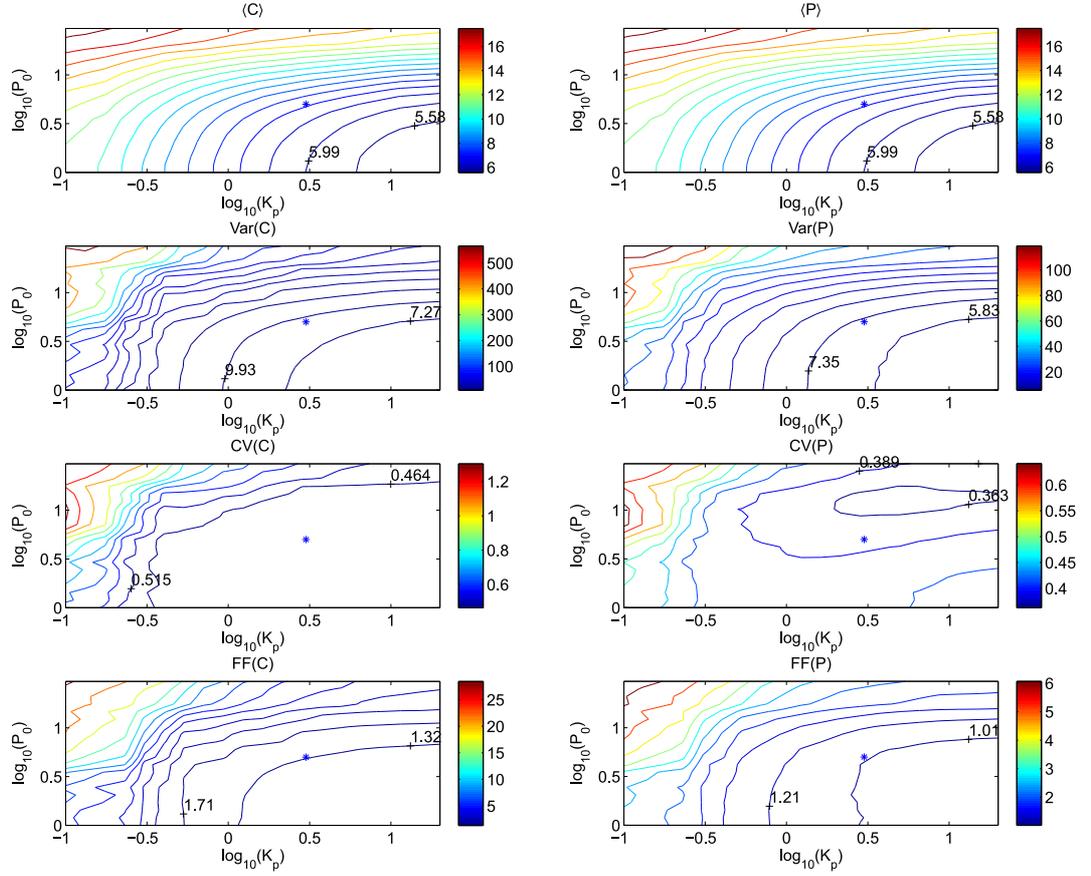}
  \end{center}
  \caption{Changes in $C$ and $P$ statistics generated by a scan over
    feedback parameters ($K_p$ and $C_0$) under feedback from $P$. It
    is interesting to note that while the behavior of the $C$ and $P$
    means is almost identical to the case of substrate feedback shown
    above, the noise in $C$ (both in terms of CV and Fano Factor) is
    significantly increased in the present case. On the other hand,
    noise in $P$ does not seem decreased in comparison to the case of
    substrate feedback. In other words, and contrary to the substrate
    feedback scenario, the behavior of substrate fluctuations is not
    reflected in product fluctuations. This is perhaps due to a
    frequency shift in substrate fluctuations, that can no longer be
    transmitted to $P$ (note that $P$ acts as low-pass filter for
    upstream fluctuations). The feedback parameter set used in the
    greatest part of the paper ($K_p=3$, $C_0=5$) is denoted by an
    asterisk.}
  \label{fig:feedback_scan_Pfeed}
\end{figure}


\clearpage
\section{Optimization over the feedback parameters using moment closure}

Due to the presence of different time scales in the substrate and enzyme dynamics, achieving
good accuracy in the calculation of $h$ is hard and thus solving
the optimization problem of Section 3.3 for obtaining the best feedback parametrization is a computationally intensive
problem. In order to get some idea of the optimal solution
in a computationally more tractable setting, we turned to the simple
moment closure method devised in \cite{master_moment}. This method,
however, requires increasing order derivatives of the reaction rates
in general, and of the feedback term in particular.

For this purpose it is therefore preferable to work with a smooth approximation of the feedback
term in the form of a Hill function,
\begin{align}
  f(x) &= \frac{kx^n}{x^n+a^n}.
\end{align}
Results in the Hill parameter space $(k,n,a)$ can then be transferred
back into the piecewise linear form $(K,x_{0})$ of the main manuscript
through e.g. a nonlinear least-squares procedure.

\begin{figure}[H]
  \begin{center}
    \includegraphics{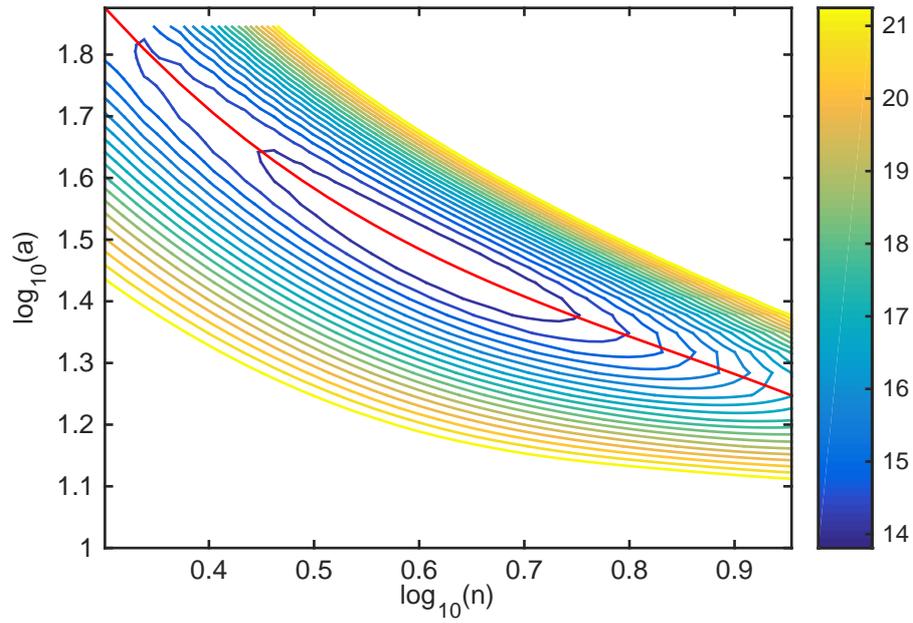}
  \end{center}
  \caption{Objective function defined in \eqref{eq:energy_Mom} for the system of moments with a Hill feedback term (assuming feedback from $C$) and moment closure using moments up to order 4. The red
  line traces the points along which $h$ is varies the least.}
  \label{fig:energy_Mom}
\end{figure}

The parameter $k$ determines the asymptote of $f$ as $x \to \infty$
and therefore only weakly affects the dynamics in a properly regulated
system where large values of $x$ are avoided. To simplify the original problem, we therefore
determined a suitable fixed value of $k$ and considered the reduced
problem
\begin{align}
  \label{eq:energy_Mom}
  \underset{n, a}\min~~ \tilde{h}(n,a):= \br{(C-C_{target})^{2}} = (\br{C}_{ss}-C_{target})^2+\mbox{Var}(C)_{ss},
\end{align}
where all moments are now computed from the closed moment equations. The function defined \eqref{eq:energy_Mom} was optimized very efficiently using the
derivative-free Nelder-Mead simplex algorithm
\cite{nelder_mead}, and also evaluated on a grid in the feedback parameter space, as shown on Fig. \ref{fig:energy_Mom}. The optimal values $n=4.15$
and $a=29.5$ were obtained for $k$ fixed at 160. It can be seen that the objective function
varies very little along the red curve; however, intermediate values of $n$ and $a$ seem to be slightly better according to the moment system.

\begin{figure}[H]
  \begin{center}
    \includegraphics[width=0.8\columnwidth]{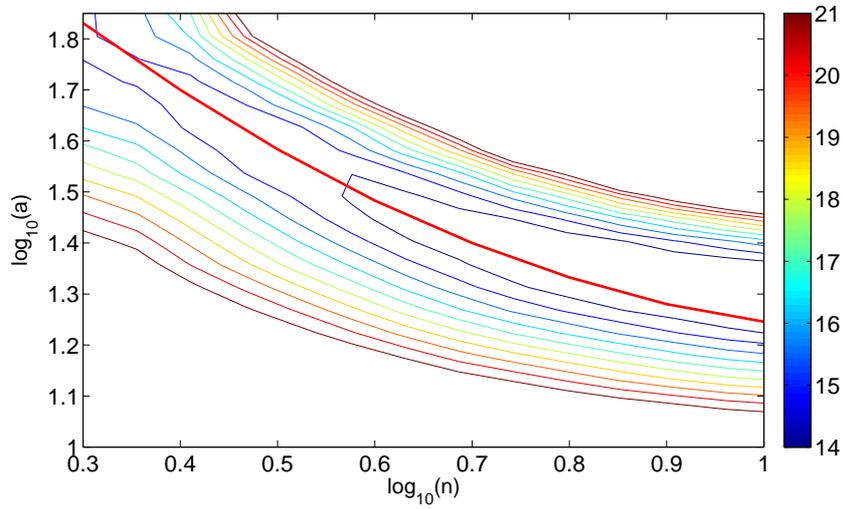}
  \end{center}
  \caption{Objective function defined in \eqref{eq:energy_Mom} computed
    by stochastic simulation, assuming feedback from $C$. The red curve from Fig. \ref{fig:energy_SSA} is overlaid.}
  \label{fig:energy_SSA}
\end{figure}

The contours of the same objective function \eqref{eq:energy_Mom}, computed with respect to the original stochastic
dynamics, is shown on Fig. \ref{fig:energy_SSA}. As the overlay of the red curve from Fig. \ref{fig:energy_Mom} indicates, this feature
is not an artifact of moment closure, but is rather visible in the SSA-based evaluation of the function. The greatest
difference from the moment closure result, is that the function now seems to get slightly smaller as $n$ increases. In this respect,
the moment closure result can serve as a good initial approximation of the optimal Hill function parameters.

The SSA result reproduces our observation made in Fig. 11 of the main text, as the optimal Hill parametrization results in a step-like function, with
very high $n$. For the range of values tested here, the optimal $n$ was found to be around 23 (the upper limit of the search interval), while $a$ was around 18.5.
These results agree very well with the results from Fig. 11, where the optimal gain was found to be equal to 30 (again, the upper limit of the search interval) and $C_0$ around 16.6, which
is very close to the ``knee'' of the Hill curve with $n=23$, $a=18.5$ and $k=160$.

\clearpage
\section{How feedback exchanges high gain with robustness}

\begin{figure}[h!tb]
\centering
  \includegraphics[width=0.8\textwidth]{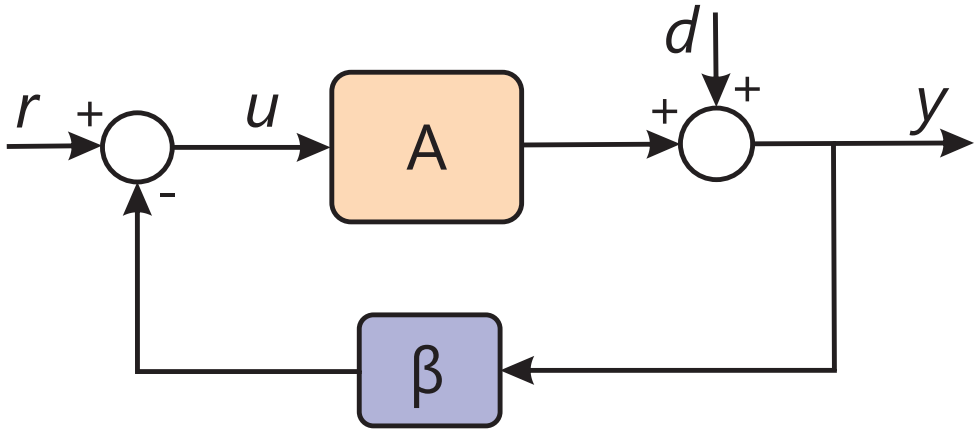}
  \caption{A simple feedback loop.}
  \label{fbloop}
\end{figure}

To gain some intuition about the role of high gain in the robustness
of feedback system, we consider here a very simple example shown on
Fig. \ref{fbloop}. The system to be regulated consists of an amplifier
A, which is simply model as a gain; that is, when $\beta=0$ and $d=0$,
the \emph{output} $y$ of A is connected to its \emph{input} $u$ by
$y=Au$, where $A$ is the gain of the amplifier. It is also possible
that an unwanted signal $d$, the \emph{disturbance}, corrupts output
of A, in which case $y=Au+d$. Assume further that $A$ is very high
($A\gg 1$) but also not known precisely and even fluctuating in
time. In this case, a given \emph{reference input} $r$ will be
translated into an output $y$ which inherits the uncertainty in the
amplifier gain. Consequently, the output of this so-called
\emph{open-loop} system (obtained for $\beta=0$) can be severely
affected by changes in $A$ and disturbance inputs $d$.

Let us now consider the \emph{closed-loop} system, obtained for
$\beta>0$. In this case, the output $y$ is multiplied by the
\emph{feedback gain} $\beta$ (which, contrary to $A$, is assumed to be
\emph{precisely} known), subtracted from $r$ and fed back into A. This is typical case of
\emph{negative feedback} because the (scaled) output is subtracted
from the input. We can now write the output as
\[y=Ae+d=A(r-\beta y)+d,\]
which implies that
\[y=\frac{A}{1+A\beta}u+\frac{1}{1+A\beta}d.\]

In this case, the closed-loop system gain from $r$ to $y$ has been
reduced from $A$ to $A(1+A\beta)^{-1}$. If $A\beta \gg 1$, this ratio
is approximately equal to $\beta^{-1}$. In other words, the gain of
the closed-loop system is now specified by the feedback gain $\beta$,
which is precisely known. The uncertainty in $A$ no longer influences
the input-output relation, while the effect of the disturbance is also
reduced by a factor of $(1+A\beta)^{-1}$.


\clearpage

\renewcommand\bibnumfmt[1]{[S #1]}

\end{document}
\fi
